\documentclass[12pt]{article}
\newcommand{\be}{\begin{equation}}
\newcommand{\ee}{\end{equation}}
\newcommand{\bi}{\begin{itemize}}
\newcommand{\ei}{\end{itemize}}
\newcommand{\bea}{\begin{eqnarray}}
\newcommand{\eea}{\end{eqnarray}}
\newcommand{\nn}{\nonumber}
\usepackage[left=2.50cm, right=2.50cm, top=2.50cm, bottom=2.50cm]{geometry}
\usepackage[utf8]{inputenc}
\usepackage{cancel}
\usepackage{placeins}
\usepackage{bm}
\usepackage[english]{babel}
\usepackage{physics}
\usepackage{amsthm}
\usepackage{mathrsfs}
\usepackage{mathtools}
\usepackage{graphicx}
\usepackage{cite}
\usepackage{changepage}
\usepackage{amssymb,amsmath}
\usepackage{verbatim}
\usepackage{empheq}
\usepackage{xcolor}
\usepackage{tikz}
\usepackage{pgfplots}
\pgfplotsset{compat=1.18}
\usepackage{float}
\usepackage{multirow}
\usepackage{multicol}
\usepackage{centernot}
\usepackage[hang,flushmargin]{footmisc}
\usepackage[normalem]{ulem}
\usepackage{hyperref}
\usetikzlibrary{tikzmark}
\usepackage{caption}
\numberwithin{equation}{section}
\hypersetup{hidelinks}

\newlength{\bibitemsep}
\newlength{\bibparskip}
\let\oldthebibliography\thebibliography
\renewcommand\thebibliography[1]{%
  \oldthebibliography{#1}%
  \setlength{\parskip}{\bibitemsep}%
  \setlength{\itemsep}{\bibparskip}%
}
\newcommand{\vectA}{\bm A}
\newcommand{\vectF}{\bm F_2}
\newcommand{\matB}{\mathsf B}
\newcommand{\matE}{\mathsf E}
\newcommand{\matJ}{\mathsf J}
\newcommand{\matG}{\mathsf G}
\newcommand{\sig}{\sigma}
\newcommand{\Rcal}{\mathcal R}

\begin{document}
\thispagestyle{empty}

\begin{minipage}{0.92\textwidth}
{\LARGE\bfseries
Edge physics and the Casimir interaction in Maxwell--Chern--Simons theory on a strip\par}

\vspace{1.0em}
{\large\bfseries Nicola Maggiore$^{a,b}$\textsuperscript{,}\footnotemark\par}

\vspace{0.6em}
{\small
$^{a}$ Dipartimento di Fisica, Universit\`a di Genova, Via Dodecaneso 33, I-16146 Genova, Italy\par
$^{b}$ Istituto Nazionale di Fisica Nucleare -- Sezione di Genova, Via Dodecaneso 33, I-16146 Genova, Italy\par
}
\end{minipage}

\footnotetext{\texttt{\href{mailto:nicola.maggiore@ge.infn.it}{nicola.maggiore@ge.infn.it}}}

\vspace{0.6em}

\begin{center}
\begin{minipage}{0.92\textwidth}
\noindent{\bfseries Abstract}\;

We study how boundaries affect Maxwell--Chern--Simons theory, a three-dimensional gauge theory that combines ordinary electromagnetic propagation with a topological Chern--Simons term. On a strip, the two boundaries can support edge excitations while the single massive bulk mode mediates a Casimir interaction between them. We use Symanzik's local boundary-field-theory framework, deriving the boundary conditions from the most general quadratic local boundary action considered here rather than imposing them by hand. Requiring these conditions to act consistently on the unique physical bulk mode selects a continuous family of admissible boundaries, characterized by an impedance and an edge velocity. The associated conserved currents form two boundary current algebras with opposite levels, and for a symmetric strip the edge modes propagate in opposite directions. The bulk and residual edge sectors factorize, leaving a single physical scattering channel for the Casimir problem. We derive its reflection amplitude, identify a stable pole-free domain, and show that the force is attractive there. In the Maxwell limit the usual long-range one-channel Casimir interaction is recovered, whereas the topological mass produces exponential screening at large separation. Outside the pole-free domain, localized surface modes can appear and must be included separately. The analysis provides a unified description of edge dynamics, boundary conditions and vacuum forces in a topologically massive gauge theory.

\end{minipage}
\end{center}

\vspace*{0pt plus 1fill}
\vspace{\baselineskip}

\newpage

\section{Introduction}
\label{intro}

Three-dimensional gauge theories provide a particularly transparent setting in
which topological and propagating sectors coexist.  Pure Chern--Simons (CS)
theory has no local bulk degrees of freedom, but on a manifold with boundary it
supports physical edge excitations whose current algebra is fixed by the CS
coupling and by boundary orientation
\cite{Witten:1988hf,Elitzur:1989nr,Moore:1989yh,Bertolini:2021iku}.  Adding a Maxwell term
produces Maxwell--Chern--Simons (MCS) theory, with one massive bulk helicity and a
topological mass scale \cite{Deser:1981wh,Deser:1982vy,Dunne:1998qy}.  On a strip
$\mathbb R^{1,1}\times[0,h]$, the width $h$ is then a genuine physical parameter:
the propagating mode can travel between the two boundaries and generate a finite
vacuum interaction.

The strip simultaneously probes three logically distinct structures.  First,
the variational principle determines which local boundary equations follow from
a chosen boundary functional.  Second, residual gauge transformations with
nonzero boundary value generate edge charges and their current algebra.  Third,
the unique gauge-invariant bulk helicity is reflected by the two boundaries and
produces the separation-dependent determinant.

These structures are often discussed with the same boundary symbols, which can
obscure their different normalizations.  The source $J^i$ couples to the
potential $A_i$; the conserved current contains both $A_i$ and the normal Maxwell
flux; the scattering amplitude acts on the dual field strength.  A consistent
calculation must preserve these distinctions when passing from Ward identities
to commutators and from boundary equations to the mode determinant.
Keeping these three layers
separate is essential.  The pullback of the gauge potential, the conserved MCS
boundary current, and the physical bulk polarization are related, but they are
not the same object.

Our starting point is the Symanzik formulation of local quantum field theory with
boundaries
\cite{Symanzik:1981wd,Vassilevich:2003xt,Bertolini:2020hgr,Amoretti:2014iza}.  Boundary
interactions are encoded by local terms, while the boundary equations are derived
from the action rather than imposed independently.

This point of view is complementary to the standard spectral formulation of a
field theory on a bounded domain.  In the latter, one starts from a differential
operator and selects a self-adjoint extension, or an elliptic boundary-value
problem, by prescribing boundary conditions directly
\cite{Vassilevich:2003xt,Asorey:2004kk,Bonneau:1999zq}.  In the Symanzik
construction, by contrast, the local boundary functional is part of the action,
and the boundary equations are its Euler--Lagrange equations.  The two
formulations agree only after one has identified the part of the Symanzik
parameter space that defines a well-posed physical spectral problem.  In a gauge
theory this second step is nontrivial because the boundary equations act on the
potential, whereas the reduced bulk phase space contains fewer physical
polarizations.

Locality also controls renormalization.  Ultraviolet divergences supported near
one boundary can be absorbed into local operators on that boundary, independently
of the couplings on the other component.  The finite interaction is the part
that cannot be assigned to either isolated boundary and therefore depends on the
round trip across the strip.  This separation is especially useful here: it
allows us to discuss the local current algebra and the nonlocal Casimir energy in
the same action without identifying them.
In the present work we first
write the complete translation-invariant quadratic functional of the tangential
pullback $A_i$ with at most one tangential derivative.  A term containing
$A_iF_2{}^i$ is retained for comparison but treated as a separate variational
branch, because its variation contains $\delta F_{2i}$ and therefore changes the
set of independent boundary data.  Within the standard tangential branch, the
zero-derivative response is a general symmetric $2\times2$ matrix, not only the
Lorentz-invariant combination $A_iA^i$.

MCS theory has one local propagating degree of freedom.  A pair of tangential
boundary equations does not automatically define a reflection map on this
one-dimensional physical polarization space: the two equations must give the
same reflected amplitude.

The compatibility condition is not an additional boundary equation imposed on
the field.  It is a restriction on the boundary couplings: it guarantees that
the two variational equations have a common one-dimensional solution space when
restricted to the physical MCS helicity.  Boundary functionals outside this
subspace can still define a formal gauge-fixed boundary-value problem, but then a
longitudinal completion must be retained explicitly and the reduced one-channel
scattering formula does not follow.  We keep this distinction visible throughout
the paper instead of referring to all local boundary couplings as equally
admissible for the physical determinant.
We impose this compatibility for all tangential
momenta.  It eliminates the one-derivative boundary response and fixes the
determinant of the symmetric boundary matrix, while leaving a continuous
two-parameter family on each boundary.  The resulting parameters have direct
physical meanings: one controls the physical reflection amplitude and the other
is the velocity of the conserved edge current.

The boundary algebra is derived from the differentiable gauge generator.  The
conserved current is
\begin{equation}
 j^i_{(\alpha)}=\sig_\alpha
 \left(\frac{1}{g^2}F^{2i}-\kappa\epsilon^{ij}A_j\right)_{\Sigma_\alpha},
\end{equation}
which includes the Maxwell normal-flux improvement.  Its central term is
nevertheless fixed solely by the CS contribution.  This reproduces the familiar
separation between the topological normalization of the current algebra, anomaly inflow, and the
dynamical propagation data
\cite{Callan:1984sa,Maggiore:2018bxr,Blasi:2010gw}.  The two strip
boundaries carry opposite levels; on the flip-symmetric branch their velocities
are equal and opposite, as in the counter-propagating configuration of Hall edge
physics \cite{Wen:1990se,Wen:1992vi,Wen:1995qn,Stone:1990iw}.

For the Casimir problem we work directly on the reduced gauge-invariant phase
space.

This choice avoids two common ambiguities.  First, the two tangential components
of the gauge potential must not be counted as two independent oscillators.
Second, a residual gauge transformation with nonzero boundary value is physical
as an edge transformation, but it is not a massive bulk polarization.  Its
boundary dynamics can contribute to the self-energy of an individual edge, yet
it does not by itself furnish a factor of the form $e^{-Qh}$ connecting the two
boundaries.  The exact factorization derived below makes both statements
algebraic rather than heuristic.
An exact factorization of the on-shell boundary operator separates the
single massive helicity from the residual pure-gauge edge sector.  The latter is
local to each boundary and contributes only $h$-independent factors to the vacuum
functional.  The interaction determinant therefore contains one physical bulk
channel, rather than a determinant over the two tangential components of
$A_i$.  This point is also required by the Maxwell limit, which in $2+1$
dimensions contains one physical degree of freedom.

The reflection amplitude is derived for a general compatible boundary.

The general compatible family remains local and continuous, despite the
one-channel reduction.  The determinant condition on the boundary matrix does
not fix its entries individually.  In particular, the edge velocity is not
forced to be the speed of light.  The flip symmetry relates the two independent
matrices only after the full spatial transformation that preserves the CS
orientation is specified.
We then
specialize to the flip-symmetric positive-impedance branch, where the round trip
is real.  Euclidean contractivity and the absence of normalizable single-boundary
surface poles are separate requirements.  In the pole-free domain the logarithmic
scattering formula is complete and the force is attractive.  Outside that
domain, the continuum determinant must be supplemented by the discrete surface
spectrum.  The resulting distinction is the precise version of the analyticity
condition needed by the scattering representation
\cite{Milton:2001yy,Bordag:2009zz,Bordag:2008gj,Rahi:2009hm,Lambrecht:2006few}.

For pure CS theory on a strip, the two boundaries support chiral sectors with
opposite current-algebra levels, while the absence of a propagating bulk mode
precludes a finite-width Casimir force \cite{Bertolini:2026qit}.  Casimir
interactions in MCS theory between parallel lines have previously been studied
for prescribed ideal boundary conditions, including perfectly conducting and
conducting/permeable configurations \cite{Milton:1990cs,Alves:2010mcs}.
Existing single-boundary MCS analyses establish the current algebra and the role
of the Maxwell improvement \cite{Maggiore:2018bxr,Blasi:2010gw}.  Here the
boundary equations are instead derived from a local Symanzik functional, so the
edge current algebra and the finite-width physical scattering problem are
organized within the same boundary parameter space.

The paper is organized as follows.  Section~\ref{geo} introduces the geometry,
conventions, sources and bulk action.  Section~\ref{bd} constructs the local
boundary functional and derives the boundary equations.  The functional Ward
identity and the canonical boundary algebra are discussed in
Sec.~\ref{ward}, while Sec.~\ref{edge} gives the scalar edge description and the
velocity constraints.  Section~\ref{maxwell} performs the physical-mode
reduction, derives the compatible boundary family and obtains the reflection
amplitudes and secular equation.  The finite interaction determinant and its
analyticity domain are developed in Sec.~\ref{casimir}.  Numerical examples and
asymptotic limits are collected in Sec.~\ref{short-h}, followed by the discussion
and outlook.  Technical derivations are given in the appendices.

\section{Geometry, conventions and bulk actions}
\label{geo}

We work in flat Minkowski $2+1$ spacetime, with coordinates
$x^\mu=(x^0,x^1,x^2)$ and metric
\be
\eta_{\mu\nu}=\mathrm{diag}(-,+,+).
\label{metric}
\ee
The Levi--Civita symbol obeys $\epsilon^{012}=+1$.  Boundary indices are
$i,j=0,1$, and we set $\epsilon^{2ij}\equiv\epsilon^{ij}$ with
$\epsilon^{01}=+1$.  The strip and its boundary are
\be
\mathcal M=\mathbb R^{1,1}\times[0,h],
\qquad
\partial\mathcal M=\Sigma_0\cup\Sigma_h.
\label{strip_geometry}
\ee
The outward normals are
\be
n_\mu\big|_{\Sigma_0}=-\delta_\mu^{\,2},
\qquad
n_\mu\big|_{\Sigma_h}=+\delta_\mu^{\,2}.
\label{normals}
\ee
It is convenient to introduce the orientation factors
\be
\sig_0=-1,
\qquad
\sig_h=+1,
\qquad
n_\mu\big|_{\Sigma_\alpha}=\sig_\alpha\delta_\mu^{\,2}.
\label{sigma}
\ee
Tangential indices are raised and lowered with
$\eta_{ij}=\mathrm{diag}(-,+)$.

Points on the two components may be represented by the embeddings
$X^\mu(x^i)=(x^0,x^1,0)$ and
$\bar X^\mu(x^i)=(x^0,x^1,h)$.  We use the same symbol for a bulk field and its
pullback when no confusion is possible.  The opposite signs in
Eq.~\eqref{sigma} are geometric and must not be absorbed into a redefinition of
the boundary couplings: they are responsible for the opposite central terms of
the two current algebras.

The abelian field strength and bulk action are
\be
F_{\mu\nu}=\partial_\mu A_\nu-\partial_\nu A_\mu,
\ee
\be
S_{\rm bulk}[A]=\int_{\mathcal M}d^3x\left[
-\frac{1}{4g^2}F_{\mu\nu}F^{\mu\nu}
+\frac{\kappa}{2}\epsilon^{\mu\nu\rho}A_\mu\partial_\nu A_\rho
\right].
\label{bulk}
\ee
We take $\kappa>0$ in the explicit mode formulas.  Reversing its sign reverses
the bulk helicity and the edge chiralities, without changing the number of
physical bulk modes.  The topological mass is
\be
m\equiv\kappa g^2>0.
\label{mdef_master}
\ee

With the normalization in Eq.~\eqref{bulk}, $[A_\mu]=1/2$, $[\kappa]=1$, and
$g^2$ is dimensionless.  The quantity $m^{-1}$ is therefore the only intrinsic
bulk length in the MCS regime.  The sign of $\kappa$ fixes the helicity, while the
Euclidean decay constant and the Casimir energy depend on $m^2$ and are invariant
under the simultaneous parity transformation that reverses $\kappa$.

The bulk equation is
\be
\frac{1}{g^2}\partial_\mu F^{\mu\nu}
+\kappa\epsilon^{\nu\mu\rho}\partial_\mu A_\rho=0.
\label{bulk_eom}
\ee
The limiting regimes are the Maxwell limit $\kappa\to0$ at fixed $g^2$, the
pure-CS limit $g^2\to\infty$ at fixed $\kappa$, and the generic MCS regime with
one massive bulk helicity of mass $m$ \cite{Deser:1981wh,Deser:1982vy}.

These limits do not commute with arbitrary choices of boundary couplings.  In
the compatible MCS family derived below the zero-derivative boundary matrix is
proportional to $\kappa$ and therefore vanishes in the Maxwell limit at fixed
dimensionless parameters.  A pure-Maxwell problem with a fixed nonzero Robin
coupling is a different path in the enlarged boundary parameter space.  We keep
this distinction explicit when comparing limiting formulas.

For perturbative quantization we may add the covariant gauge-fixing term
\be
S_{\rm gf}[A]=-\frac{1}{2\xi}\int_{\mathcal M}d^3x\,
(\partial_\mu A^\mu)^2.
\label{gf}
\ee
Its boundary variation is proportional to $\delta A^2$ and does not alter the
tangential equations derived below.  In the abelian theory the Faddeev--Popov
operator is field independent.  We do not evaluate a gauge-fixed vector/ghost
determinant: the scattering calculation is performed directly on the reduced
gauge-invariant phase space, where the physical spectrum is independent of
$\xi$.

Independent sources for the tangential pullback are introduced on the two
boundary components,
\be
S_J=\sum_{\alpha=0,h}\int_{\Sigma_\alpha}d^2x\,
J^i_{(\alpha)}A_i.
\label{sources}
\ee
They generate correlators of $A_i$,
\be
\frac{\delta W}{\delta J^i_{(\alpha)}(x)}
=\left\langle A_i(x)\right\rangle.
\label{boundary_field_generator}
\ee
This quantity is the boundary value of the gauge potential, not the conserved
MCS current.  The latter will be identified from the Ward identity and the
Gauss generator.

Higher functional derivatives generate connected correlators of the boundary
pullback.  Since the sources on $\Sigma_0$ and $\Sigma_h$ are independent,
separability is implemented before finite-width propagation is included.  The
only coupling between the two components in the quadratic theory comes from
solving the bulk equations across the strip.  This is the field-theory version
of the multiple-reflection construction used later.

After Wick rotation, tangential Euclidean momenta are denoted by $(\zeta,k)$ and
\be
Q(\zeta,k)=\sqrt{\zeta^2+k^2+m^2}.
\label{Qdef_master}
\ee
The physical evanescent profiles are proportional to $e^{\pm Qx^2}$.

\section{Local boundary action and derived boundary conditions}
\label{bd}

Following Symanzik, local boundary interactions are written as
\be
S_{\rm bd}=\int_{\mathcal M}d^3x\left[
\delta(x^2)\mathcal L_0+\delta(x^2-h)\mathcal L_h\right].
\label{Sbd}
\ee
We assume translation invariance along each boundary, work at quadratic order,
and retain at most one tangential derivative.  Introduce the column vectors and
the antisymmetric matrix
\be
\vectA=\begin{pmatrix}A_0\\A_1\end{pmatrix},
\qquad
\vectF=\begin{pmatrix}F^{20}\\F^{21}\end{pmatrix},
\qquad
\matE=\begin{pmatrix}0&1\\-1&0\end{pmatrix}.
\label{boundary_vectors}
\ee
Within the standard tangential variational problem, the complete local density
in this truncation is
\begin{align}
\mathcal L_\alpha^{\rm tan}
={}&\frac12\vectA^{\rm T}\matB_{(\alpha)}\vectA
+\frac{d_0^{(\alpha)}}2\vectA^{\rm T}\matE\,\partial_0\vectA
+\frac{d_1^{(\alpha)}}2\vectA^{\rm T}\matE\,\partial_1\vectA,
\label{Lbd_tan}\\
\matB_{(\alpha)}={}&
\begin{pmatrix}
b_{00}^{(\alpha)}&b_{01}^{(\alpha)}\\
b_{01}^{(\alpha)}&b_{11}^{(\alpha)}
\end{pmatrix}.
\label{Bmatrix}
\end{align}
The symmetric matrix contains the three independent zero-derivative couplings.
The restricted Lorentz-invariant choice $a_1A_iA^i/2$ is recovered only for
$b_{00}=-a_1$, $b_{01}=0$, and $b_{11}=a_1$.

No boundary Lorentz invariance is assumed.  This is important because a material
or effective boundary may select a preferred rest frame even when the bulk is
relativistic.  The off-diagonal coefficient $b_{01}$ and the independent values
of $b_{00}$ and $b_{11}$ are therefore part of the local Symanzik data.  Their
presence is also what allows a continuous edge velocity without forcing
$|v_\alpha|=1$.  After tangential integration by parts, the derivative tensor is
necessarily antisymmetric in its two field labels and therefore has the form
shown in Eq.~\eqref{Lbd_tan}.

To see this, write the coefficient of $A_i\partial_aA_j$ as
$C^{ija}$.  The symmetric part $C^{(ij)a}$ differs from a total derivative only
by a derivative of the constant coefficient, which vanishes.  The remaining
antisymmetric tensor has one component for each derivative direction and is
proportional to $\epsilon^{ij}$.  No use of the bulk equations is made in this
classification.  The coefficients $b_{ij}$ have mass dimension one, whereas
$d_0$ and $d_1$ are dimensionless.

Terms with two tangential derivatives have dimension three in the present power
counting and lie outside the truncation.  Terms proportional to a tangential
total derivative do not affect the variational equations.  We also exclude
bilocal expressions coupling $\Sigma_0$ directly to $\Sigma_h$: such a term
would violate Symanzik separability and would insert an inter-edge interaction by
hand instead of deriving it from bulk propagation.

For comparison with the broader local basis, we also keep
\be
\mathcal L_\alpha^{\perp}=a_4^{(\alpha)}A_iF_2{}^i.
\label{Lbd_normal}
\ee
This term belongs to a different variational class: its variation contains
$A_i\delta F_2{}^i$ and hence treats a normal derivative as independent boundary
data.  We shall distinguish this branch explicitly rather than count
$a_4^{(\alpha)}$ as another entry of the tangential matrix
$\matB_{(\alpha)}$.

Varying the Maxwell and CS bulk terms gives
\begin{align}
\delta S_{\rm Max}\big|_{\partial\mathcal M}
&=-\sum_{\alpha=0,h}\frac{\sig_\alpha}{g^2}
\int_{\Sigma_\alpha}d^2x\,F^{2i}\delta A_i,
\label{varMax_boundary}\\
\delta S_{\rm CS}\big|_{\partial\mathcal M}
&=\sum_{\alpha=0,h}\sig_\alpha\frac\kappa2
\int_{\Sigma_\alpha}d^2x\,\epsilon^{ij}A_j\delta A_i.
\label{varCS_boundary}
\end{align}
The signs follow directly from the outward normals in Eq.~\eqref{normals}.  On
the tangential branch $a_4^{(\alpha)}=0$, the two boundary equations can be
written compactly as
\be
-\frac{\sig_\alpha}{g^2}\vectF
+\left[\matB_{(\alpha)}+\varrho_\alpha(\partial)\matE\right]\vectA=0,
\qquad
\varrho_\alpha(\partial)=\sig_\alpha\frac\kappa2
+d_0^{(\alpha)}\partial_0+d_1^{(\alpha)}\partial_1.
\label{BC_general}
\ee
In components,
\begin{align}
0={}&-\frac{\sig_\alpha}{g^2}F^{20}
+b_{00}^{(\alpha)}A_0
+\left[b_{01}^{(\alpha)}+\varrho_\alpha(\partial)\right]A_1,
\label{BC0_exp}\\
0={}&-\frac{\sig_\alpha}{g^2}F^{21}
+\left[b_{01}^{(\alpha)}-\varrho_\alpha(\partial)\right]A_0
+b_{11}^{(\alpha)}A_1.
\label{BC1_exp}
\end{align}
These are the variational boundary equations before any restriction to the
physical MCS polarization.

For generic values of the coefficients, Eqs.~\eqref{BC0_exp} and
\eqref{BC1_exp} are two independent linear relations among the tangential
potential and the normal flux.  Their derivation does not yet guarantee that the
associated gauge-fixed operator is elliptic, that the physical MCS helicity is
closed under reflection, or that the Euclidean round trip is analytic.  These
are separate questions, addressed successively in Secs.~\ref{maxwell} and
\ref{casimir}.  This layered analysis prevents variational admissibility from
being conflated with spectral stability.

\subsection{Boundary branches and the edge-active sector}
\label{branches}

The variation of Eq.~\eqref{Lbd_normal} contains
\be
\delta S_{\rm bd}^{\perp}\supset
\sum_{\alpha=0,h}\int_{\Sigma_\alpha}d^2x\,
 a_4^{(\alpha)}A_i\,\delta F_2{}^i.
\label{deltaF_variation}
\ee
If $F_{2i}$ is included among the independently varied boundary data, stationarity
requires
\be
 a_4^{(\alpha)}A_i\big|_{\Sigma_\alpha}=0.
\label{a4_branch_condition}
\ee
The condition in Eq.~\eqref{a4_branch_condition} is only the coefficient of
$\delta F_{2i}$.  The coefficient of $\delta A_i$ must be imposed at the same
time.  If $a_4^{(\alpha)}\neq0$, the first condition gives $A_i=0$ on the
boundary; its tangential derivatives then vanish there, and the remaining
$\delta A_i$ equation reduces to
\be
\left(a_4^{(\alpha)}-\frac{\sig_\alpha}{g^2}\right)
F^{2i}\big|_{\Sigma_\alpha}=0.
\label{a4_flux_condition}
\ee
Thus the normal-field-strength term separates the variational problem into the
following cases.

\begin{table}[!t]
\centering
\small
\begin{tabular}{|c|c|c|c|}
\hline
Branch on $\Sigma_\alpha$ & Independent data & Boundary conditions & Role below\\ \hline
$a_4^{(\alpha)}=0$ & $\delta A_i$ & Eqs.~\eqref{BC0_exp}--\eqref{BC1_exp} & edge-active tangential branch\\ \hline
$a_4^{(\alpha)}=\sig_\alpha/g^2$ & $\delta A_i,\delta F_{2i}$ & $A_i=0$ & tuned Dirichlet branch\\ \hline
$a_4^{(\alpha)}\notin\{0,\sig_\alpha/g^2\}$ & $\delta A_i,\delta F_{2i}$ & $A_i=0$, $F^{2i}=0$ & generically overconstrained\\ \hline
\end{tabular}
\caption{Variational branches associated with the normal-field-strength term.}
\label{tab:BC_branches}
\end{table}
\FloatBarrier

The tuned Dirichlet branch permits only residual gauge transformations whose
boundary value is constant.  It therefore carries no local edge algebra of the
type studied in Secs.~\ref{ward} and~\ref{edge}.  For generic nonzero
$a_4^{(\alpha)}$ the simultaneous Dirichlet and normal-flux conditions impose
both boundary data of the physical second-order normal problem and are
generically overconstraining; this case is not used below.

The classification is local to each boundary component.  At the level of the
variational problem one may, for example, choose the edge-active branch on one
side and the tuned Dirichlet branch on the other.  The corresponding complete
gauge-fixed spectral problem also requires the normal/gauge boundary condition
and is outside the reduced edge-active problem studied here.  We use the
edge-active choice on both sides.

More precisely, on the tuned Dirichlet branch $A_i=0$ forces
$\partial_i\lambda=0$ for a residual gauge transformation preserving the
boundary data.  Only a boundary-constant zero mode survives, so there is no
local Kac--Moody density.  This branch belongs to a different variational class
and will not be mixed with the edge-active scattering family developed below.
From now on, the current-algebra and Casimir analysis refers to
\be
 a_4^{(0)}=a_4^{(h)}=0.
\label{edge_active_branch}
\ee
The general tangential equations remain Eqs.~\eqref{BC_general}--\eqref{BC1_exp}.
Compatibility with the single physical helicity will impose an additional,
independent restriction in Sec.~\ref{maxwell}.

\section{Broken Ward identity and boundary current algebra}
\label{ward}

A gauge transformation $\delta_\lambda A_\mu=\partial_\mu\lambda$ leaves the
Maxwell term invariant.  The CS action changes by a boundary term, while the
variation of $S_{\rm bd}^{\rm tan}$ is the tangential divergence of its boundary
Euler derivative.  Defining
\be
\bm{\mathcal E}_{(\alpha)}
=\matB_{(\alpha)}\vectA
+d_0^{(\alpha)}\matE\,\partial_0\vectA
+d_1^{(\alpha)}\matE\,\partial_1\vectA,
\label{boundary_Euler_vector}
\ee
the local combination selected by the gauge variation is
\be
\bm j_{(\alpha)}
=\bm{\mathcal E}_{(\alpha)}
-\sig_\alpha\frac\kappa2\matE\vectA.
\label{current_from_Ward}
\ee
The functional Ward identity in the presence of the sources
Eq.~\eqref{sources} has the schematic local form
\be
\partial_iJ^i_{(\alpha)}
+\partial_i\left\langle j^i_{(\alpha)}\right\rangle
=\mathcal W^{(\alpha)}_{\rm gf}.
\label{Ward_W}
\ee

The identity may equivalently be obtained by changing integration variables in
the functional integral, in the standard derivation of functional Ward
identities \cite{ZinnJustin,PeskinSchroeder}.  The source term produces
$-\int\lambda\partial_iJ^i$, the boundary functional gives
$-\int\lambda\partial_i\mathcal E^i$, and the CS variation supplies the
orientation-dependent curl.  Their sum is the divergence of
Eq.~\eqref{current_from_Ward}.  Contact terms generated by differentiating
Eq.~\eqref{Ward_W} are local and depend on the chosen source normalization; they
should not be used to identify $A_i$ itself with the physical current.

Here $\mathcal W_{\rm gf}$ is the standard gauge-fixing insertion.  It vanishes
for residual transformations preserving the gauge condition and does not alter
the equal-time central term.  At zero sources the current is conserved.

The same current follows directly from the normal bulk equation.  The $\nu=2$
component of Eq.~\eqref{bulk_eom} is
\be
\partial_i\left(
\frac1{g^2}F^{2i}-\kappa\epsilon^{ij}A_j
\right)=0.
\label{normal_eom}
\ee
With the intrinsic orientation of each boundary component, define
\be
j^i_{(\alpha)}
=\sig_\alpha\left(
\frac1{g^2}F^{2i}-\kappa\epsilon^{ij}A_j
\right)_{\Sigma_\alpha},
\qquad
\partial_i j^i_{(\alpha)}=0.
\label{physical_current}
\ee
Using the variational equations, Eq.~\eqref{physical_current} is exactly
Eq.~\eqref{current_from_Ward}.  In particular,
\begin{align}
j^0_{(\alpha)}={}&
 b_{00}^{(\alpha)}A_0
+\left(b_{01}^{(\alpha)}-\sig_\alpha\frac\kappa2\right)A_1
+d_0^{(\alpha)}\partial_0A_1+d_1^{(\alpha)}\partial_1A_1,
\label{current0_general}\\
j^1_{(\alpha)}={}&
 \left(b_{01}^{(\alpha)}+\sig_\alpha\frac\kappa2\right)A_0
+b_{11}^{(\alpha)}A_1
-d_0^{(\alpha)}\partial_0A_0-d_1^{(\alpha)}\partial_1A_0.
\label{current1_general}
\end{align}
Equations~\eqref{boundary_field_generator} and~\eqref{physical_current} make the
normalization distinction explicit: $\delta W/\delta J^i$ generates the boundary
value of $A_i$, whereas the conserved current is the flux combination
Eq.~\eqref{physical_current}.

This distinction also resolves the apparent dimensional paradox that would arise
from assigning a central term proportional to $\kappa$ directly to the field
$A_i$.  Since $[A_i]=1/2$, an algebra for $A_i$ itself would carry a coefficient
with inverse mass dimension.  The current $j^0$, however, contains an explicit
factor of $\kappa$ (or the equivalent Maxwell flux through the boundary
equation), has the correct dimension, and acquires a central term proportional
to $\kappa$.

The physical scattering family derived in Sec.~\ref{compatibility_subsection}
has $d_0^{(\alpha)}=d_1^{(\alpha)}=0$.  In particular, its boundary functional
contains no independent time-derivative term and therefore adds no boundary
contribution to the symplectic form.  The canonical derivation below refers to
this compatible family.  If the unreduced branch with $d_0^{(\alpha)}\neq0$ is
studied instead, its boundary symplectic term must be included in the charge
analysis.

The central extension on the compatible family is most cleanly derived from the
differentiable gauge generator.  A direct BJL extraction from time-ordered
correlators of $A_i$ \cite{Bjorken:1966jh,Weinberg:QTF1} would instead give the
algebra of the boundary gauge field in the source normalization.  Its coefficient is inverse in $\kappa$, as required by
dimensions.  Multiplying by the CS normalization entering the physical charge
converts that field algebra into the current algebra proportional to $\kappa$.
The canonical generator performs this normalization automatically and is
therefore less prone to ambiguity.

On a constant-$x^0$ slice, the canonical momentum and Gauss constraint are
\be
\pi^a=\frac1{g^2}F_{0a}+\frac\kappa2\epsilon^{ab}A_b,
\qquad a,b=1,2,
\label{canonical_momentum}
\ee
\be
\mathcal G=\partial_a\pi^a
+\frac\kappa2\epsilon^{ab}\partial_aA_b\approx0.
\label{gauss_constraint}
\ee
Allowing the gauge parameter to be nonzero on the boundary, the differentiable
generator is
\be
G[\lambda]=-
\int_{\Sigma_t}d^2x\,\lambda\mathcal G
+\sum_{\alpha=0,h}Q_\alpha[\lambda],
\label{gauge_generator}
\ee
with
\be
Q_\alpha[\lambda]
=\int_{\Sigma_\alpha\cap\Sigma_t}dx^1\,\lambda\rho_\alpha,
\qquad
\rho_\alpha=\sig_\alpha\left(\pi^2-\frac\kappa2A_1\right)
=j^0_{(\alpha)}.
\label{boundary_charge}
\ee
The generator gives
\be
\delta_\eta\pi^2=-\frac\kappa2\partial_1\eta,
\qquad
\delta_\eta\rho_\alpha=-\sig_\alpha\kappa\partial_1\eta.
\label{charge_variation}
\ee
Consequently,
\be
\{Q_\alpha[\lambda],Q_\beta[\eta]\}
=-\delta_{\alpha\beta}\sig_\alpha\kappa
\int dx^1\,\lambda\partial_1\eta,
\label{charge_algebra}
\ee

No boundary equation has been used in obtaining the central term.  The result is
a property of the presymplectic structure and the differentiability of the gauge
generator.  The boundary equations enter only when the charge density is
expressed in terms of the boundary values of $A_0$ and $A_1$, as in
Eqs.~\eqref{current0_param} and~\eqref{current_chirality}.  Upon quantization,
Eq.~\eqref{charge_algebra} gives
\be
[\rho_\alpha(x^1),\rho_\beta(y^1)]
=-i\delta_{\alpha\beta}\sig_\alpha\kappa
\partial_{x^1}\delta(x^1-y^1).
\label{KM_density}
\ee
The local levels are therefore
\be
k_\alpha=-\sig_\alpha\kappa,
\qquad
k_0=-k_h.
\label{KM_levels}
\ee
The Maxwell term is present in the current through $F_{02}$ but is gauge
invariant and does not contribute to the central extension.

The result is therefore insensitive to the zero-derivative entries of
$\matB_{(\alpha)}$ on the compatible family.  Those entries determine which
linear combination of $A_0$ and $A_1$ represents the charge density on shell,
but the transformation of that density under a boundary gauge parameter is
fixed by the CS part of the canonical momentum.  This is the precise sense in
which the level is topological while the current profile is dynamical.  Since
the two boundary supports are disjoint, mixed local brackets vanish.  A global
integrated Gauss constraint may relate zero modes, but it does not change either
local algebra.

\section{Edge scalar fields and chiral dynamics}
\label{edge}

The physical current is conserved independently on each connected boundary
component.  Locally it can therefore be written as
\be
j^i_{(0)}=\epsilon^{ij}\partial_j\phi,
\qquad
j^i_{(h)}=\epsilon^{ij}\partial_j\bar\phi.
\label{scalar_potentials}
\ee
The scalar potentials are defined up to boundary zero modes.  This relation does
not identify $A_i$ with $\partial_i\phi$; it parametrizes the conserved current
Eq.~\eqref{physical_current}.

The local parametrization is sufficient for the nonzero-momentum algebra.  The
constant modes are constrained by the integrated Gauss law and may correlate the
total charges on the two components.  Such a global relation does not modify the
local commutators or the continuum Casimir determinant, which is built from
nonzero-frequency bulk oscillators.

\subsection{Compatibility family and current velocity}
\label{edge_scalars_subsect}

The physical reflection problem derived in Sec.~\ref{maxwell} selects
\be
 d_0^{(\alpha)}=d_1^{(\alpha)}=0,
\qquad
\det\matB_{(\alpha)}=-\frac{\kappa^2}{4}.
\label{compatibility_preview}
\ee
On the finite-density component of this family, write
\be
\matB_{(\alpha)}=\kappa
\begin{pmatrix}
\gamma_\alpha&\gamma_\alpha v_\alpha-\sig_\alpha/2\\
\gamma_\alpha v_\alpha-\sig_\alpha/2&
\gamma_\alpha v_\alpha^2-\sig_\alpha v_\alpha
\end{pmatrix}.
\label{compatible_B_edge}
\ee
Substituting into Eqs.~\eqref{current0_general}--\eqref{current1_general} gives
\be
j^0_{(\alpha)}=\kappa\left[
\gamma_\alpha A_0+
(\gamma_\alpha v_\alpha-\sig_\alpha)A_1
\right],
\label{current0_param}
\ee
\be
j^1_{(\alpha)}=v_\alpha j^0_{(\alpha)}.
\label{current_chirality}
\ee
Thus $v_\alpha$ is the propagation velocity of the conserved edge current.  In
terms of the scalar potentials,
\be
(\partial_0+v_0\partial_1)\phi=0,
\qquad
(\partial_0+v_h\partial_1)\bar\phi=0.
\label{chiral_bosons}
\ee
The current algebra becomes
\be
[\partial_1\phi_\alpha(x^1),\partial_1\phi_\beta(y^1)]
=-i\delta_{\alpha\beta}\sig_\alpha\kappa
\partial_{x^1}\delta(x^1-y^1),
\label{scalar_algebra}
\ee
where $\phi_0\equiv\phi$ and $\phi_h\equiv\bar\phi$.

A first-order scalar action reproducing the same local algebra and equation of
motion is, in the standard chiral-boson representation of a boundary current
algebra \cite{Maggiore:2017vjf},
\be
S_{\rm edge}^{(\alpha)}
=\frac{1}{2k_\alpha}\int d^2x\,
\partial_1\phi_\alpha
\left(\partial_0\phi_\alpha+v_\alpha\partial_1\phi_\alpha\right),
\qquad
k_\alpha=-\sig_\alpha\kappa.
\label{edge_scalar_action}
\ee
Up to zero modes and total derivatives, its second-class constraint yields
Eq.~\eqref{scalar_algebra}.  The corresponding quadratic Hamiltonian is
\be
H_\alpha=-\frac{v_\alpha}{2k_\alpha}
\int dx^1\,\rho_\alpha^2,
\qquad
k_\alpha=-\sig_\alpha\kappa.
\label{edge_Hamiltonian}
\ee
Indeed, with
$[\rho(x),\rho(y)]=ik_\alpha\partial_x\delta(x-y)$, the Heisenberg equation
gives $\partial_0\rho=-v_\alpha\partial_1\rho$.  Since
$\rho=\partial_1\phi_\alpha$, this is equivalent to the chiral equation
modulo a spatially constant mode.  The action is not postulated as an additional
boundary degree of freedom: it is the local effective representation of the
residual gauge sector already present in the MCS variational problem.

For $\kappa>0$, positivity requires
\be
\sig_\alpha v_\alpha>0.
\label{edge_positivity}
\ee
This is independent of the compatibility condition and of the Euclidean
analyticity conditions imposed later.

A microscopic relativistic completion may further require
$|v_\alpha|\leq1$.  We do not impose that optional causal bound in the formal
classification, because an effective boundary medium can break boundary Lorentz
invariance.  The numerical examples used below satisfy it.

\subsection{Flip symmetry}
\label{edge_velocities}

A reflection of $x^2$ alone reverses the three-dimensional orientation and would
change the sign of the CS term.  The symmetry at fixed $\kappa$ is instead the
orientation-preserving spatial flip
\be
\mathsf F:\quad
(x^0,x^1,x^2)\longmapsto(x^0,-x^1,h-x^2),
\label{flip_coordinates}
\ee
with
\be
A_0\longmapsto A_0,
\qquad
A_1\longmapsto-A_1,
\qquad
A_2\longmapsto-A_2.
\label{flip_fields}
\ee
It exchanges the two boundaries and implies
\be
b_{00}^{(h)}=b_{00}^{(0)},
\qquad
b_{01}^{(h)}=-b_{01}^{(0)},
\qquad
b_{11}^{(h)}=b_{11}^{(0)}.
\label{flip_B}
\ee
Within Eq.~\eqref{compatible_B_edge},
\be
\gamma_h=\gamma_0\equiv\gamma,
\qquad
v_h=-v_0.
\label{flip_parameters}
\ee
We denote $v_0\equiv v$, so that edge positivity selects
\be
v<0,
\qquad
v_h=-v>0.
\label{flip_edge_positive}
\ee
The two currents counter-propagate, while their local levels remain opposite and
fixed by $\kappa$.

The flip acts simultaneously on geometry and field components.  Writing only
$x^2\mapsto h-x^2$ would be a parity transformation and would reverse the sign
of the CS density.  The additional reflection of $x^1$ in
Eq.~\eqref{flip_coordinates} is therefore essential when the same value of
$\kappa$ is used on both sides of the strip.

\section{Maxwell--Chern--Simons theory on a strip: spectral setup}
\label{maxwell}

\subsection{Gauge-invariant physical reduction}
\label{physical_reduction}

Introduce the dual field strength
\be
f^\mu=\frac12\epsilon^{\mu\nu\rho}F_{\nu\rho}.
\label{dual_field}
\ee
The MCS equation is equivalent to
\be
\epsilon^{\mu\nu\rho}\partial_\nu f_\rho+m f^\mu=0,
\qquad
\partial_\mu f^\mu=0,
\qquad
(\Box-m^2)f^\mu=0.
\label{self_dual_equations}
\ee
This first-order system carries one local bulk polarization.

The second-order Klein--Gordon equation alone would contain two formal
polarizations, but the first-order helicity equation projects onto one of them.
The physical reduction must therefore be performed before constructing the
boundary determinant.  Counting the two tangential entries of $A_i$ as separate
reflection channels would double the Maxwell limit and is not compatible with
the reduced phase space.  For $m>0$, any on-shell potential can be decomposed
locally as
\be
A_\mu=A_\mu^{\rm phys}+\partial_\mu\lambda,
\qquad
A_\mu^{\rm phys}=-\frac1m f_\mu.
\label{on_shell_decomposition}
\ee
The difference has vanishing field strength and is therefore pure gauge on the
simply connected strip.

For a plane wave
\be
A^\mu=e^\mu(p)e^{-i\omega x^0+ikx^1+ipx^2},
\qquad
\omega^2=k^2+p^2+m^2,
\label{plane_wave}
\ee
a convenient physical polarization is
\be
e^\mu(p)=\mathcal N
\left(
\frac{-ikm+p\omega}{m^2+p^2},
\frac{kp-im\omega}{m^2+p^2},
1
\right).
\label{physical_polarization}
\ee
It obeys
\be
p_\mu e^\mu=0,
\qquad
i\epsilon^{\mu\nu\rho}p_\nu e_\rho=-m e^\mu.
\label{helicity_equations}
\ee

The normalization $\mathcal N$ cancels from every reflection coefficient.  The
apparent singularity of Eq.~\eqref{physical_polarization} at
$m^2+p^2=0$ is a coordinate singularity of this polarization chart.  Whenever a
surface branch crosses that locus, the first-order field equation must be solved
directly; this check is carried out in Appendix~\ref{secular_app}.

Thus the incident and reflected physical waves carry one amplitude each.

\subsection{Compatibility of the tangential equations}
\label{compatibility_subsection}

In momentum space,
\be
\varrho_\alpha(\omega,k)=\sig_\alpha\frac\kappa2
-id_0^{(\alpha)}\omega+id_1^{(\alpha)}k.
\label{rho_momentum}
\ee
Let $V_{0,\alpha}(p)$ and $V_{1,\alpha}(p)$ be the two boundary expressions
Eqs.~\eqref{BC0_exp}--\eqref{BC1_exp} evaluated on
Eq.~\eqref{physical_polarization}.  Both equations define the same reflected
amplitude only if
\be
\Delta_\alpha
=V_{0,\alpha}(p)V_{1,\alpha}(-p)
-V_{1,\alpha}(p)V_{0,\alpha}(-p)=0.
\label{compatibility_def}
\ee
Direct substitution gives
\be
\Delta_\alpha=\mathcal K_\alpha(p)
\left\{
\det\matB_{(\alpha)}+
[\varrho_\alpha(\omega,k)-\sig_\alpha\kappa]^2
\right\},
\label{compatibility_factor}
\ee
where $\mathcal K_\alpha(p)$ is nonzero for a generic physical wave.  Requiring
compatibility for all tangential momenta yields
\be
 d_0^{(\alpha)}=d_1^{(\alpha)}=0,
\qquad
\det\matB_{(\alpha)}=-\frac{\kappa^2}{4}.
\label{compatibility_conditions}
\ee
The one-derivative terms remain legitimate local boundary interactions in the
unreduced gauge-fixed problem, but they do not close on the one-helicity physical
scattering subspace for all momenta.  The zero-derivative compatible family is
continuous: one condition is imposed on the three entries of the symmetric
matrix.

The compatibility equation is polynomial in the tangential momenta.  Its
coefficients of $\omega^2$, $k^2$, and $\omega k$ force the two derivative
couplings to vanish; no low-momentum approximation is involved.  The remaining
constant condition is the determinant constraint.  Thus the continuous velocity
found below is not an infrared artifact and does not require discarding the
normal flux.

In the restricted Lorentz-invariant subspace
$b_{00}=-a_1$, $b_{01}=0$, $b_{11}=a_1$, the determinant condition would force
$a_1^2=\kappa^2/4$.  Treating $a_1$ as arbitrary in that restricted subspace
therefore overconstrains the physical reflection problem.  The general symmetric
matrix resolves this: it retains two continuous parameters while satisfying the
same determinant condition.

For $b_{00}^{(\alpha)}\neq0$, the family is parametrized by
Eq.~\eqref{compatible_B_edge}.

Conversely, given a compatible matrix with $b_{00}\neq0$, one recovers
\be
\gamma_\alpha=\frac{b_{00}^{(\alpha)}}{\kappa},
\qquad
v_\alpha=\frac{b_{01}^{(\alpha)}+\sig_\alpha\kappa/2}
{b_{00}^{(\alpha)}}.
\label{inverse_parameterization}
\ee
The remaining entry is then fixed by the determinant.  These relations show
that $v_\alpha$ is not introduced by hand: it is an invariant ratio of local
boundary couplings on the finite-density chart.  The chart extends continuously
to $\gamma_\alpha=0$ with
$b_{01}^{(\alpha)}=-\sig_\alpha\kappa/2$.  The complementary component
$b_{00}^{(\alpha)}=0$, $b_{01}^{(\alpha)}=+\sig_\alpha\kappa/2$ has
$j^0_{(\alpha)}\equiv0$ and is not part of the finite-density branch considered
below.

The one-channel reduction can be established without choosing a gauge.  Define
\be
\matJ_{(\alpha)}=\matB_{(\alpha)}-
\sig_\alpha\frac\kappa2\matE,
\qquad
\matG_{(\alpha)}=\matB_{(\alpha)}+
\sig_\alpha\frac\kappa2\matE,
\label{JG_matrices}
\ee
and
\be
u_\alpha=\begin{pmatrix}1\\v_\alpha\end{pmatrix},
\qquad
w_\alpha=\begin{pmatrix}\gamma_\alpha\\
\gamma_\alpha v_\alpha-\sig_\alpha\end{pmatrix}.
\label{uw_vectors}
\ee
On the compatible family,
\be
\matJ_{(\alpha)}=\kappa u_\alpha w_\alpha^{\rm T},
\qquad
\matG_{(\alpha)}=\kappa w_\alpha u_\alpha^{\rm T},
\qquad
\det(u_\alpha,w_\alpha)=-\sig_\alpha.
\label{JG_factorization}
\ee
Using Eq.~\eqref{on_shell_decomposition}, the full boundary equation becomes
\be
u_\alpha\big(w_\alpha^{\rm T}\vectA^{\rm phys}\big)
+w_\alpha\big(u_\alpha^{\rm T}\partial\lambda\big)=0.
\label{decoupled_boundary_equation}
\ee
The two vectors are linearly independent, so this is equivalent to
\be
w_\alpha^{\rm T}\vectA^{\rm phys}=0,
\qquad
(\partial_0+v_\alpha\partial_1)\lambda=0.
\label{physical_gauge_split}
\ee

The first equation is a single physical boundary condition, as required for a
second-order scalar-like normal problem.  The second is the chiral residual-gauge
condition already obtained from current conservation.  Their simultaneous
appearance is not a gauge choice: it follows from the rank-one factorization of
the variational boundary operator.

A longitudinal amplitude cannot cancel a physical one at the boundary.  The
massive bulk helicity and the residual edge sector factorize exactly.  Gauge
transformations vanishing at both boundaries remove the interior interpolation
of $\lambda$; the two surviving boundary values obey local chiral equations and
do not introduce a propagation factor depending on $h$.

\subsection{Boundary conditions in momentum space and reflection amplitudes}
\label{reflection_subsection}

After Wick rotation $\omega=i\zeta$, let
\be
Q=\sqrt{\zeta^2+k^2+m^2}.
\label{Qdef}
\ee
The convention for the incident evanescent momentum at $\Sigma_\alpha$ is
$p=\sig_\alpha iQ$.  Substitution of Eq.~\eqref{physical_polarization} into the
physical boundary condition in Eq.~\eqref{physical_gauge_split} gives
\be
r_\alpha(\zeta,k)
=\frac{QX_\alpha+mY_\alpha}{QX_\alpha-mY_\alpha},
\label{general_r}
\ee
with
\begin{align}
X_\alpha&=\sig_\alpha\gamma_\alpha\zeta
+ik(\sig_\alpha\gamma_\alpha v_\alpha-1),
\label{Xdef}\\
Y_\alpha&=(\gamma_\alpha v_\alpha-\sig_\alpha)\zeta
+i\gamma_\alpha k.
\label{Ydef}
\end{align}
The two original tangential equations give the same ratio precisely because of
Eq.~\eqref{compatibility_conditions}.

For real $\omega,k,p$, the numerator entering the reflection coefficient has the
form $iA+pB$ with $A$ and $B$ real.  Hence the outgoing and incoming numerators
are complex conjugates up to a sign and $|r_\alpha|=1$.  The local boundary
functional is conservative.  Possible instabilities are instead detected by
poles on the physical sheet or by a failure of the Euclidean contour
continuation.

The physical scalar nature of the result can also be seen directly from
$\psi=f^2$.  The tangential components of the self-dual field satisfy
\begin{align}
m f_0+\partial_2f_1&=\partial_1\psi,
\\
\partial_2f_0+m f_1&=\partial_0\psi.
\label{dual_component_relations}
\end{align}
Away from the exceptional tangential light cone, combining these equations with
$w_\alpha^{\rm T}\vectA^{\rm phys}=0$ and
$A_i^{\rm phys}=-f_i/m$ gives the oblique scalar condition
\begin{align}
0={}&m\left[
\gamma_\alpha\partial_1+
(\gamma_\alpha v_\alpha-\sig_\alpha)\partial_0
\right]\psi
\nn\\
&-\partial_2\left[
\gamma_\alpha\partial_0+
(\gamma_\alpha v_\alpha-\sig_\alpha)\partial_1
\right]\psi.
\label{scalar_oblique_BC}
\end{align}
This derivation shows explicitly that the boundary operator is first order in the
normal derivative even though it contains tangential derivatives.  The
exceptional locus is treated directly with the first-order helicity equation in
Appendix~\ref{secular_app}.

In Euclidean signature the exceptional locus is absent away from the origin, so
Eq.~\eqref{scalar_oblique_BC} also gives an independent derivation of
Eq.~\eqref{general_r}.  It exhibits the physical problem as one massive scalar
amplitude subject to an oblique boundary operator containing tangential
frequency and momentum.  Only on special parameter lines does this operator
reduce to an angle-independent Robin-like coefficient.  The generic compatible
MCS boundary is therefore more general than a momentum-independent scalar Robin
condition.

For unrelated boundaries the physical round trip is
\be
M(\zeta,k;h)=r_0(\zeta,k)r_h(\zeta,k)e^{-2Qh}.
\label{round_trip_scalar}
\ee

The factor $e^{-Qh}$ appears once for propagation from $\Sigma_0$ to
$\Sigma_h$ and once on the return path.  A pole of an individual $r_\alpha$ is a
single-boundary state, whereas a zero of $D=1-M$ is a finite-strip normal mode.
These two types of spectral singularity should not be identified.

The real-frequency secular equation is the analytic continuation of
\be
D(\zeta,k;h)=1-M(\zeta,k;h)=0.
\label{secular_equation}
\ee

The use of a scalar secular function does not assume flip symmetry.  For two
unrelated compatible boundaries, $r_0r_h$ can be complex pointwise on the
Euclidean plane.  For real boundary parameters Eq.~\eqref{general_r} obeys
\be
r_\alpha(\zeta,-k)=r_\alpha(\zeta,k)^*,
\qquad
r_\alpha(-\zeta,k)=r_\alpha(\zeta,k)^*,
\label{general_reality_symmetry}
\ee
whereas simultaneous inversion of both tangential momenta leaves
$r_\alpha$ unchanged.  Consequently
$D(\zeta,-k;h)=D(\zeta,k;h)^*$ (and likewise under $\zeta\to-\zeta$).
If the contour is free of singularities and the logarithm is continued from
$h=\infty$, the paired contributions are complex conjugates and the integrated
vacuum energy is real.  The flip-symmetric family is singled out because it
makes reality and positivity pointwise through Eq.~\eqref{conjugate_r}.

\subsection{Flip-symmetric reflectivity and physical poles}
\label{strip_limits_subsection}

For the flip relations Eq.~\eqref{flip_parameters},
\be
r_h(\zeta,k)=r_0(\zeta,k)^*.
\label{conjugate_r}
\ee
Let
\be
c\equiv1+\gamma v.
\label{cdef}
\ee
The round-trip reflectivity is
\begin{align}
\Rcal(\zeta,k)=|r_0|^2
=\frac{
 k^2(Qc-m\gamma)^2
+\zeta^2(Q\gamma-mc)^2}
{
 k^2(Qc+m\gamma)^2
+\zeta^2(Q\gamma+mc)^2}.
\label{reflectivity}
\end{align}
The denominator minus the numerator equals
\be
4Qm\,\gamma c\,(\zeta^2+k^2).
\label{reflectivity_difference}
\ee
On the positive-impedance branch $\gamma>0$, strict Euclidean contractivity at
nonzero tangential momentum requires $c>0$ and gives
\be
0\leq\Rcal(\zeta,k)<1.
\label{Rrange}
\ee
The threshold value is direction independent,
\be
\Rcal_*=\lim_{\zeta,k\to0}\Rcal
=\left(\frac{c-\gamma}{c+\gamma}\right)^2.
\label{threshold_R}
\ee

The angular dependence away from threshold is a genuine consequence of the
oblique boundary operator.  It disappears on the line $c=\gamma$, where the
same scalar reflection factor is obtained for every direction in the Euclidean
$(\zeta,k)$ plane.  At large momentum $Q\gg m$, all pole-free members approach
perfect reflection, explaining the universal short-distance limit.

On the threshold-transparent line
\be
c=\gamma,
\label{transparent_line}
\ee
one has
\be
\Rcal(\zeta,k)=\left(\frac{Q-m}{Q+m}\right)^2.
\label{transparent_R}
\ee
For real propagating normal momentum, the boundary action is conservative and
$|r_\alpha|=1$.

This real-axis unitarity follows before flip symmetry is imposed.  Flip symmetry
is used only to make the Euclidean round trip equal to the nonnegative quantity
$\Rcal=|r_0|^2$.  The physical interpretation is therefore the same as for a
lossless mirror: the real-frequency wave is reflected with a phase, while the
Wick-rotated amplitude controls convergence of the vacuum multiple-reflection
series.  Contractivity is therefore a property of the Wick-rotated amplitude,
not a statement of dissipation.  The same scattering logic underlies finite
Casimir constructions based directly on boundary data \cite{Blasi:1992mm}.

Contractivity does not by itself exclude a physical surface mode.  At the lower
boundary the physical condition reduces to
\be
\gamma A_0+cA_1=0.
\label{lower_physical_bc}
\ee
For a mode $e^{-\lambda x^2}$, $\lambda>0$, the helicity equation gives
\be
\lambda(ck-\gamma\omega)+m(\gamma k-c\omega)=0,
\qquad
\omega^2=k^2+m^2-\lambda^2.
\label{surface_pole_equation}
\ee
If $\gamma>c>0$, there is a normalizable branch
\be
\omega_{\rm s}(k)=\frac{ck+m\sqrt{\gamma^2-c^2}}{\gamma},
\qquad
\lambda_{\rm s}(k)=
\frac{k\sqrt{\gamma^2-c^2}-cm}{\gamma}.
\label{surface_branch}
\ee
It exists for $k>cm/\sqrt{\gamma^2-c^2}$.  At its onset $\lambda=0$, it
joins the bulk continuum; at larger momentum it becomes exponentially localized.
In a finite strip, the two single-boundary branches can hybridize, and the
resulting level splitting is $h$ dependent.  The branch therefore contributes a
discrete strip spectrum and must not be omitted from a contour calculation.
This is why the continuum logarithm alone is incomplete in that region even
though it remains formally contractive on the Euclidean axis.  If
$c\geq\gamma>0$, no finite normalizable pole remains on the physical sheet.
We therefore define the
positive-impedance, edge-positive, pole-free flip-symmetric domain by
\be
v<0,
\qquad
0<\gamma\leq c=1+\gamma v.
\label{pole_free_domain}
\ee
The closure $\gamma=0$ is perfectly reflecting and pole free.

We restrict the explicit Casimir analysis to this positive-impedance component.
Other sign components of the determinant surface can be parametrized, but the
simultaneous requirements of edge-energy positivity, Euclidean reality and pole
placement must then be reconsidered.  No conclusion about their force is implied
by the results below.  The crossing of the surface branch through the
polarization-chart singularity and the limiting case $c=\gamma$ are checked
directly in Appendix~\ref{secular_app}.

\section{Finite Casimir energy on a strip}
\label{casimir}

\subsection{Finite interaction and one-channel mode count}
\label{finite_logic}

The vacuum energy contains bulk-local, boundary-local and interaction pieces.

One may organize the same separation diagrammatically.  Vacuum graphs that never
visit both boundaries renormalize the bulk or one isolated boundary.  Graphs that
contain at least one propagation from each boundary to the other are finite at
nonzero $h$ after the local subtractions and are resummed by the logarithm of the
round trip.  At quadratic order this graphical classification is equivalent to
the determinant subtraction.

The interaction is defined by subtracting the infinite-separation limit,
\be
E_{\rm int}(h)=E_{\rm vac}(h)-E_{\rm vac}(\infty).
\label{interaction_subtraction}
\ee
Local ultraviolet terms are removed by bulk and single-boundary counterterms.
The remaining $h$ dependence is nonlocal and finite.

In heat-kernel language, the divergent coefficients are integrals of local bulk
invariants and local invariants on each connected boundary component.  They do
not contain the nonlocal factor connecting $\Sigma_0$ and $\Sigma_h$.  The
subtraction in Eq.~\eqref{interaction_subtraction} therefore removes the
scheme-dependent self-energies without altering the measurable force.  This is
the same subtraction used in standard Lifshitz and multiple-scattering
treatments of the Casimir effect
\cite{Milton:2001yy,Bordag:2001qi,Bordag:2009zz,Emig:2007cf,Rahi:2009hm,Reynaud:2010vr}.

The exact factorization in Eq.~\eqref{physical_gauge_split} fixes the reduced
physical mode count.  The residual-gauge values on $\Sigma_0$ and $\Sigma_h$
generate the two chiral edge sectors, whose nonzero-mode spectra are local to
the corresponding boundary and independent of $h$.  Their isolated-edge vacuum
terms therefore cancel in Eq.~\eqref{interaction_subtraction}.  The unique
massive helicity is the only physical oscillator carrying the propagation factor
$e^{-Qh}$.  Thus the separation-dependent physical secular determinant contains
one channel.

The statement concerns the $h$-dependent part of the determinant.  Each isolated
edge still has a chiral partition function and a local vacuum energy.  These
quantities are meaningful boundary observables, but they are identical in the
finite-strip and infinite-separation configurations and cancel from
$E_{\rm int}$.  The local Kac--Moody sectors therefore coexist with, but do not
double, the propagating bulk round trip.  The possible relation between global
zero modes in a two-boundary geometry is conceptually distinct from this local
factorization \cite{Henneaux:2019sjx}.

\subsection{Scattering/log-determinant representation}
\label{scattering_representation}

If the single-boundary amplitudes are analytic on the Wick-rotation contour, no
surface pole crosses that contour, and
\be
|r_0r_h|e^{-2Qh}<1,
\label{general_analyticity_condition}
\ee
the interaction energy per unit length is
\be
E_{\rm int}(h)=\frac12
\int_{-\infty}^{+\infty}\frac{d\zeta}{2\pi}
\int_{-\infty}^{+\infty}\frac{dk}{2\pi}
\ln\left[1-r_0(\zeta,k)r_h(\zeta,k)e^{-2Qh}\right].
\label{Eint_scattering}
\ee
The branch of the logarithm is chosen continuously from $h=\infty$, where the
round trip vanishes.  This prescription fixes the interaction unambiguously in
the pole-free domain.  The condition is stronger than the mere algebraic
existence of $r_\alpha$: a denominator may be nonzero on the Euclidean line and
still hide a physical pole crossed during the contour rotation.

The force is
\be
F(h)=-\frac{dE_{\rm int}}{dh}
=-\int\frac{d\zeta\,dk}{(2\pi)^2}
Q\frac{r_0r_h e^{-2Qh}}{1-r_0r_h e^{-2Qh}}.
\label{force_scattering}
\ee

The force formula is often numerically preferable because it avoids subtracting
nearby vacuum energies and differentiating interpolated data.  It also makes the
sign transparent when the round trip is real and nonnegative.  The normalization
is fixed by the factor $1/2$ in the zero-point energy and by the derivative of
$e^{-2Qh}$; no additional polarization multiplicity is present.

For the flip-symmetric pole-free family,
\be
E_{\rm int}(h)=\frac12\int\frac{d\zeta\,dk}{(2\pi)^2}
\ln\left[1-\Rcal(\zeta,k)e^{-2Qh}\right],
\label{flip_energy}
\ee
\be
F(h)=-\int\frac{d\zeta\,dk}{(2\pi)^2}
Q\frac{\Rcal e^{-2Qh}}{1-\Rcal e^{-2Qh}}.
\label{flip_force}
\ee
Since $0\leq\Rcal<1$ in Eq.~\eqref{pole_free_domain} and $Q>0$ for
$m>0$, one has
\be
E_{\rm int}(h)<0,
\qquad
F(h)<0.
\label{attraction}
\ee
Indeed, every term in the expansion
$\ln(1-\Rcal e^{-2Qh})=-\sum_{n\geq1}\Rcal^n e^{-2nQh}/n$ is negative.  The
same termwise argument applied to Eq.~\eqref{flip_force} makes the attraction
manifest.  In the reflection-related pole-free family this result is consistent
with general attraction theorems based on reflection positivity
\cite{Kenneth:2006vr}, although here the sign follows directly from
$\Rcal\geq0$.

The attraction statement is restricted to the pole-free flip-symmetric domain.
When a surface pole exists, its finite-strip splitting must be included before
the sign of the total force can be assessed.

\subsubsection{Relation to a scalar scattering problem}
\label{diag_check}

Equation~\eqref{Eint_scattering} has the standard form for one scalar channel,
consistently with the scalar result previously found for parallel conducting MCS
lines \cite{Milton:1990cs}, but the effective boundary equation is the oblique,
momentum-dependent condition Eq.~\eqref{scalar_oblique_BC}.  The reduction is
therefore a statement about the physical mode count, not an identification of all
compatible MCS boundaries with ordinary momentum-independent Robin data.

A conventional scalar Robin boundary has a coefficient
$r_{\rm R}(Q)=(Q-\lambda)/(Q+\lambda)$ depending only on $Q$.  The general MCS
amplitude depends separately on $\zeta$ and $k$ because the boundary matrix also
controls a chiral velocity.  The threshold-transparent line is a special case:
its round-trip reflectivity is $[(Q-m)/(Q+m)]^2$, but this simplification should
not be extrapolated to the entire compatible family.  In the Maxwell limit of
the compatible MCS family, the boundary matrix scales to zero and the dual scalar
is
perfectly reflected, as shown in Sec.~\ref{Maxwell_limit_subsection}.  Generic
Robin conditions of pure Maxwell theory belong to a different limiting family in
which a nonzero scalar boundary coupling is held fixed.

\subsubsection{Numerical variables and checks}
\label{subnumerics}

Set
\be
x=mh,
\qquad
\nu=h\sqrt{\zeta^2+k^2},
\qquad
q=\sqrt{\nu^2+x^2},
\label{dimensionless_defs}
\ee
and use $\zeta=(\nu/h)\cos\theta$, $k=(\nu/h)\sin\theta$.  Then
\begin{align}
h^2E_{\rm int}
&=\frac1{8\pi^2}\int_0^\infty\nu\,d\nu
\int_0^{2\pi}d\theta\,
\ln\left[1-\Rcal_x(\nu,\theta)e^{-2q}\right],
\label{Eint_dimensionless}\\
h^3F
&=-\frac1{4\pi^2}\int_0^\infty\nu\,d\nu
\int_0^{2\pi}d\theta\,
q\frac{\Rcal_x(\nu,\theta)e^{-2q}}
{1-\Rcal_x(\nu,\theta)e^{-2q}},
\label{Force_dimensionless}
\end{align}
where
\be
\Rcal_x(\nu,\theta)=
\frac{
\sin^2\theta\,(qc-x\gamma)^2
+\cos^2\theta\,(q\gamma-xc)^2}
{
\sin^2\theta\,(qc+x\gamma)^2
+\cos^2\theta\,(q\gamma+xc)^2}.
\label{scaled_R}
\ee
The integrands are exponentially suppressed for large $\nu$.

For each $(\nu,\theta)$ the denominator in Eq.~\eqref{scaled_R} is evaluated
before the logarithm, and the pole-free inequalities provide a direct numerical
check that the argument remains between zero and one.  The radial integration is
split into a low-momentum interval and an exponentially convergent tail.  The
angular integral is smooth and periodic.  We have compared adaptive radial
quadrature with an independent fixed high-order Gauss--Legendre rule; the values
used in the figures agree well beyond the displayed precision.  This direct
momentum-space evaluation is complementary to time-domain numerical approaches
to Casimir forces \cite{Rodriguez:2009fzr}.

A further check does not require a second integration.  Writing
\be
f(x;\gamma,v)\equiv h^2E_{\rm int}(h),
\qquad x=mh,
\label{scaling_function}
\ee
the definition $F=-dE_{\rm int}/dh$ at fixed $m$ implies the exact identity
\be
h^3F(h)=2f(x;\gamma,v)-x\,\partial_x f(x;\gamma,v).
\label{energy_force_identity}
\ee
We evaluate the force both from Eq.~\eqref{Force_dimensionless} and by applying
Eq.~\eqref{energy_force_identity} with a five-point numerical derivative of the
energy.  At representative values throughout the plotted interval, the relative
difference is below $5\times10^{-10}$.  At $x=0.2,1,3$, adaptive and fixed-order
quadratures for the energy differ by less than $10^{-14}$ in absolute units.  The
Maxwell values at $x=0$ are inserted analytically rather than inferred from a
finite-mass extrapolation.

\subsection{Zeta-function viewpoint}
\label{zeta_argument}

The same result follows from the argument principle.  For fixed tangential
momentum, let $D(\omega,k;h)$ be the secular function obtained by propagating the
single physical amplitude from one boundary to the other.  For a large box in
the $x^2$ direction, the zeros of $D$ discretize the normal spectrum.  The box
can be removed after subtracting the isolated-boundary spectra, and the
logarithmic derivative in Eq.~\eqref{zeta_contour} counts each physical zero
once, consistently with the one-dimensional polarization space.  The difference
between the finite-strip and infinite-separation spectra can be written as
\be
E_{\rm int}(h)=\frac12\int\frac{dk}{2\pi}
\oint_{\mathcal C}\frac{d\omega}{2\pi i}\,
\omega\,\partial_\omega
\ln\frac{D(\omega,k;h)}{D(\omega,k;\infty)}.
\label{zeta_contour}
\ee
In the pole-free domain the contour can be rotated to the imaginary axis without
crossing a zero or pole, giving Eq.~\eqref{Eint_scattering}.

The normalization by $D(\omega,k;\infty)$ removes factors associated with a
single reflection and leaves only the round trip.  Equivalently, one can
differentiate with respect to $h$, evaluate the convergent force contour, and
integrate back with $E_{\rm int}(\infty)=0$.  Both prescriptions agree in the
pole-free domain.  If a surface mode is present, its pole is crossed during the
deformation and the associated
residue supplies the missing discrete contribution.  Euclidean contractivity
alone is therefore not sufficient: the pole analysis in
Sec.~\ref{strip_limits_subsection} is an independent part of the spectral
problem.  The strictly massless point $Q=0$ is the familiar integrable threshold
and is understood by the standard infrared limiting prescription
\cite{Elizalde:1995hck,Kirsten2001}.

\section{Asymptotic regimes and consistency checks}
\label{short-h}

\subsection{Worked flip-symmetric example and numerics}
\label{numerics_limits_8}

We use two representative points in the domain Eq.~\eqref{pole_free_domain}:
\be
(\gamma,v)=(0.30,-1/2),
\qquad
(\gamma,v)=(1/2,-1).
\label{worked_parameters}
\ee
For the first, $c=0.85$ and $\Rcal_*=(0.55/1.15)^2$; the second lies on the
threshold-transparent line $c=\gamma=1/2$.  Both have positive edge Hamiltonians
and no physical single-boundary pole.

The first point is generic: its threshold reflectivity is nonzero and the
large-distance interaction starts with $e^{-2mh}/h$.  The second point is chosen
to isolate the effect of threshold transparency; its leading coefficient
vanishes and the decay starts at $e^{-2mh}/h^3$.  The comparison therefore tests
both asymptotic formulas without leaving the analytically controlled domain.

\begin{figure}[H]
\centering
\includegraphics[width=0.97\linewidth]{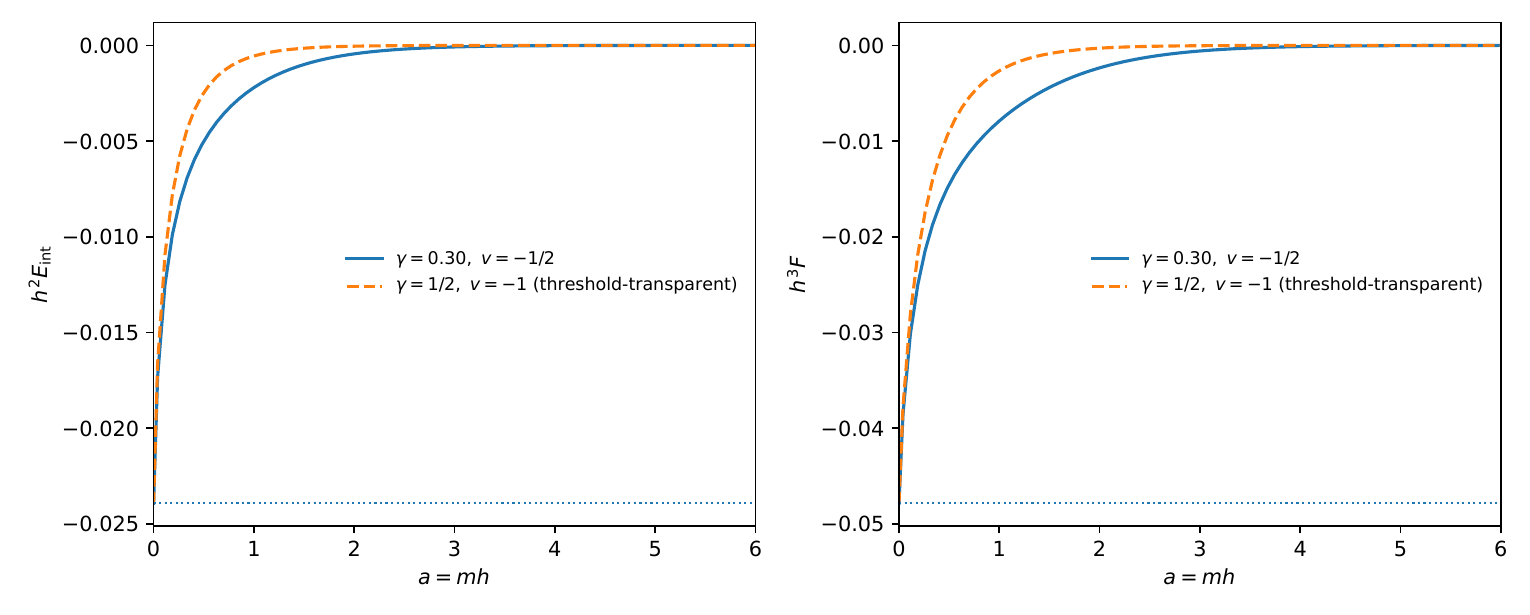}
\caption{Numerical evaluation of the pole-free flip-symmetric family.  Left:
$h^2E_{\rm int}$ as a function of $a=mh$.  Right: $h^3F$.  The solid curve is
the generic point $(\gamma,v)=(0.30,-1/2)$ and the dashed curve is the
threshold-transparent point $(1/2,-1)$.  The horizontal dotted lines are the
one-channel Maxwell limits $-\zeta(3)/(16\pi)$ and
$-\zeta(3)/(8\pi)$, respectively.}
\label{fig:worked_example}
\end{figure}

Figure~\ref{fig:worked_example} shows the rescaled energy and force.  The two
curves share the universal one-channel Maxwell plateaux
\be
h^2E_{\rm int}\longrightarrow-\frac{\zeta(3)}{16\pi},
\qquad
h^3F\longrightarrow-\frac{\zeta(3)}{8\pi}
\qquad (mh\to0).
\label{UV_plateaux}
\ee
The generic point has the Yukawa tail controlled by $\Rcal_*$, while the
threshold-transparent curve has an additional inverse-power suppression.

\begin{figure}[H]
\centering
\includegraphics[width=\linewidth]{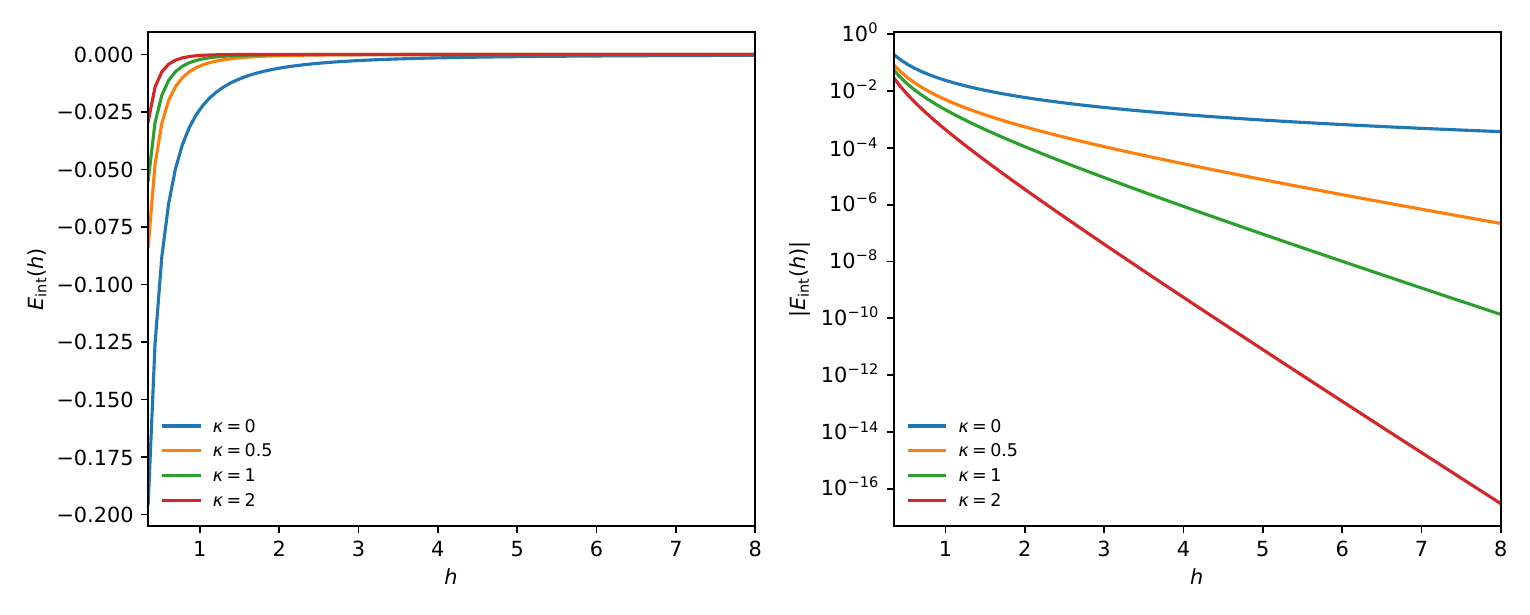}
\caption{Unrescaled interaction energy for $g^2=1$ and
$(\gamma,v)=(0.30,-1/2)$, with $\kappa=0,0.5,1,2$.  Left: $E_{\rm int}(h)$.
Right: $|E_{\rm int}(h)|$ on a logarithmic scale.  All curves have the same
short-distance coefficient $-\zeta(3)/(16\pi h^2)$; the nonzero-$\kappa$ curves
are Yukawa screened at large separation.}
\label{fig:kappa_comparison}
\end{figure}

To display the role of the topological mass directly,
Fig.~\ref{fig:kappa_comparison} shows the unrescaled interaction at $g^2=1$ and
fixed $(\gamma,v)=(0.30,-1/2)$ for several values of $\kappa$.  The Maxwell curve
remains long ranged, whereas every nonzero $\kappa$ produces exponential
screening on the scale $m^{-1}=(\kappa g^2)^{-1}$.  No sign change occurs in the
pole-free flip-symmetric domain.

At fixed $(\gamma,v)$ the dimensionless function depends only on $mh$; changing
$\kappa$ at fixed $g^2$ therefore moves the crossover scale while preserving the
short-distance coefficient.  The logarithmic panel distinguishes the Maxwell
power law from the exponentially screened MCS curves without rescaling the
energy by powers of $h$.

\subsection[Pure Chern--Simons limit at fixed kappa]{Pure Chern--Simons limit $g^2\to\infty$ at fixed $\kappa$}
\label{pureCS_8}

At fixed $\kappa$, $g^2\to\infty$ sends $m\to\infty$.

The limit must be taken after the interaction subtraction.  Local boundary
partition functions and the edge Hamiltonians do not vanish; only the part that
requires a physical bulk round trip is suppressed.  Thus the disappearance of
the force does not imply the disappearance of the edge states.  Since
$Q\geq m$, the physical propagation factor vanishes exponentially,
\be
E_{\rm int}(h)\longrightarrow0,
\qquad
F(h)\longrightarrow0.
\label{pure_CS_limit}
\ee
The edge levels and velocities remain finite.  Thus the pure-CS limit preserves
the local boundary algebra while eliminating the propagation-mediated Casimir
interaction, consistently with the absence of a local bulk degree of freedom.

\subsection[Pure Maxwell limit]{Pure Maxwell limit $\kappa=0$}
\label{Maxwell_limit_subsection}

Along the compatible MCS family, $\matB_{(\alpha)}$ is proportional to $\kappa$.
Taking $\kappa\to0$ at fixed $g^2$, $\gamma_\alpha$ and $v_\alpha$ gives
\be
m\to0,
\qquad
\matB_{(\alpha)}\to0,
\qquad
\Rcal\to1.
\label{Maxwell_limit_data}
\ee
The boundary equations become $F^{2i}=0$.  In the Maxwell dual-scalar
representation $F^{\mu\nu}\propto\epsilon^{\mu\nu\rho}\partial_\rho\varphi$,
these conditions set the tangential derivatives of $\varphi$ to zero; for
nonzero boundary momentum this is Dirichlet, up to the constant zero mode.  The
interaction is therefore
\be
E_{\rm int}(h)\big|_{m=0}
=\frac12\int\frac{d\zeta\,dk}{(2\pi)^2}
\ln\left(1-e^{-2h\sqrt{\zeta^2+k^2}}\right)
=-\frac{\zeta(3)}{16\pi h^2},
\label{Maxwell_energy}
\ee
\be
F(h)\big|_{m=0}=-\frac{\zeta(3)}{8\pi h^3}.
\label{Maxwell_force}
\ee
This is the standard result for one massless scalar channel, as required by the
single propagating degree of freedom of Maxwell theory in $2+1$ dimensions.

The factor of two relative to a two-component tangential determinant is
physical, not conventional.  After reduction by gauge symmetry, Maxwell theory
in three spacetime dimensions is dual to one scalar and has no second physical
tangential polarization.  The same counting is inherited continuously from the
massive MCS helicity.

Generic scalar Robin conditions can be imposed in pure Maxwell theory, but they
are obtained by holding an independent scalar boundary coupling fixed; they are
not the limit of Eq.~\eqref{compatible_B_edge} at fixed dimensionless
$(\gamma,v)$.

\subsection[Short-distance regime]{Short-distance regime $h\to0$}
\label{short_distance_subsection}

For $mh\ll1$, the integral is dominated by $Q\gg m$, where
$\Rcal\to1$.  Hence
\be
E_{\rm int}(h)
=-\frac{\zeta(3)}{16\pi h^2}+o(h^{-2}),
\qquad
F(h)=-\frac{\zeta(3)}{8\pi h^3}+o(h^{-3}).
\label{short_distance}
\ee
The coefficient is independent of $\gamma$ and $v$ inside the pole-free domain
and counts the single physical bulk channel.

Boundary-dependent corrections are suppressed by powers of $mh$.  Their detailed
form is generally angle dependent, but they do not modify the leading Weyl
coefficient.  The short-distance limit is therefore a useful diagnostic of the
physical mode count.

\subsection[Large-separation regime]{Large-separation regime $h\to\infty$}
\label{large_distance_subsection}

For generic pole-free data with $\Rcal_*\neq0$, the first round trip gives
\be
E_{\rm int}(h)
\sim-\Rcal_*e^{-2mh}
\left(\frac{m}{8\pi h}+\frac1{16\pi h^2}\right),
\qquad mh\gg1,
\label{generic_large_E}
\ee
\be
F(h)
\sim-\Rcal_*e^{-2mh}
\left(\frac{m^2}{4\pi h}+\frac{m}{4\pi h^2}
+\frac1{8\pi h^3}\right).
\label{generic_large_F}
\ee
Higher round trips carry $e^{-4mh},e^{-6mh},\ldots$ and are exponentially
subleading.  Corrections from the momentum dependence of $\Rcal$ add inverse
powers of $h$ to the same leading exponential.  The formulas above therefore
include the complete leading Yukawa term and its first kinematic power
correction.

On the threshold-transparent line, $\Rcal_*=0$ and
\be
E_{\rm int}(h)
\sim-\frac{e^{-2mh}}{64\pi m h^3},
\qquad
F(h)
\sim-\frac{e^{-2mh}}{64\pi h^3}
\left(2+\frac3{mh}\right).
\label{transparent_asymptotics}
\ee
The topological correlation length is $m^{-1}$ in both cases, but threshold
transparency adds two inverse powers of $h$.

The exponential is fixed by the lightest bulk excitation and is independent of
the boundary parameters.  Those parameters enter the prefactor through the
threshold reflection amplitude.  This clean separation between the screening
length and the boundary transparency is one of the main advantages of the
scattering representation.

\section{Discussion and outlook}
\label{discussion}

The strip problem separates into a variational classification, a
boundary charge algebra and a physical one-channel scattering problem.

This separation also clarifies which conclusions are robust under changes of the
boundary functional.  The existence of one massive bulk helicity and the
orientation dependence of the CS level are bulk facts.  The velocity, impedance,
threshold reflectivity and possible surface spectrum are boundary data.  A local
boundary term can modify the latter without changing the former.  The complete
zero-derivative tangential response is a symmetric matrix, while the
one-derivative terms are allowed by locality but do not preserve the reduced MCS
helicity for all momenta.  Compatibility fixes only the determinant of the
symmetric matrix, leaving a continuous family rather than a discrete set of
boundary conditions.

The algebraic and dynamical data have different origins.  The current
Eq.~\eqref{physical_current} contains the Maxwell normal flux, but its central
extension is fixed by the CS symplectic term and by orientation.  The local
levels are $k_0=-k_h$, whereas the velocities are boundary parameters.

There is no contradiction between an algebra proportional to $\kappa$ and a
physical gauge-field bracket proportional to $1/\kappa$: the conserved charge
density contains the CS normalization.  Stating the current explicitly is
therefore essential when comparing different source conventions in the
literature.  The full orientation-preserving flip relates the two matrices and
gives equal and
opposite velocities.  Positivity of the edge Hamiltonians imposes an additional
sign condition that is logically independent of the variational and spectral
constraints.

The physical determinant contains one bulk channel.

The local boundary current itself vanishes on a purely self-dual bulk wave:
$F^{20}/g^2=\kappa A_1$ and $F^{21}/g^2=-\kappa A_0$.  The nontrivial current
algebra is carried by the residual-gauge edge sector, whereas the bulk helicity
carries the finite-width propagation.  This complementary support of the two
sectors is another way to see why their determinants should not be multiplied as
independent bulk polarizations.  The one-channel result is not obtained by
forming a two-component determinant and subsequently deleting one eigenvalue.
It follows from the exact factorization
Eqs.~\eqref{JG_factorization}--\eqref{physical_gauge_split}: the massive helicity
and the residual edge gauge transformations satisfy independent boundary
equations.  The latter generate the Kac--Moody sectors but have no
separation-dependent propagation factor.  This also fixes the Maxwell
coefficient to $-\zeta(3)/(16\pi h^2)$.

The analyticity domain requires two checks.  Euclidean contractivity controls the
multiple-reflection series, while the real-frequency pole analysis determines
whether a discrete surface spectrum is crossed during Wick rotation.  For the
positive-impedance flip-symmetric branch, the residue-free result is established
for Eq.~\eqref{pole_free_domain}.  In this domain the round-trip reflectivity is
nonnegative and the force is attractive.  The region $\gamma>c>0$ supports a
physical surface branch; its finite-strip hybridization is a separate problem
and may alter the full interaction after the discrete contribution is included.

The Maxwell limit is also more specific than a generic Robin benchmark.

This qualification is relevant when comparing distinct paths in parameter
space.  Sending $\kappa$ to zero while keeping $(\gamma,v)$ fixed is not the same
operation as first setting $\kappa=0$ and then choosing an arbitrary scalar
boundary mass.  The two procedures define different families of boundary-value
problems, even though both may be described by scalar scattering amplitudes.
Along the compatible MCS family the boundary matrix vanishes with $\kappa$, and the
dual scalar is perfectly reflected.  Pure-Maxwell Robin conditions are possible,
but correspond to a different scaling of the boundary functional.  In the
opposite pure-CS limit the edge algebra survives while the Casimir force
vanishes.  The strip therefore separates topological boundary information from
propagation-mediated finite-width observables.

Several extensions are natural.

A first priority is the surface-mode region, where the residue contribution can
be obtained from the finite-strip secular equation and compared with the
continuum term.  A second is the general non-flip family, for which the product
$r_0r_h$ need not be pointwise real even though the integrated energy can remain
real.  Both problems can be addressed within the scalar physical reduction
without returning to a two-component tangential determinant.  The non-flip
compatible family can be studied
by keeping the complex scalar round trip and imposing reality and stability
directly.  The surface-pole sector requires the discrete mode splitting in
addition to the continuum determinant.  Boundary terms involving normal
derivatives, dynamical boundary matter, finite temperature and inhomogeneous
response can all enlarge the reflection space.  The same distinction between
local current algebra and propagation-mediated interaction may also be useful in
BF-type, symmetric-tensor and higher-rank gauge theories
\cite{Blasi:2019wpq,Maggiore:2019wie,Amoretti:2013xya,Bertolini:2023wie,Bertolini:2023sqa,Bertolini:2022sao,Bertolini:2024yur,Bertolini:2025qcy,Bertolini:2025jul,Bertolini:2025goo,Bertolini:2025rhz}.

\section*{Acknowledgments}
I thank Erica Bertolini and Dario Ferraro for insightful discussions. I am grateful to Alberto Blasi for introducing me to the wonders of the Casimir effect and its many subtleties.

\appendix

\section{Boundary variation and boundary conditions}
\label{BCapp}

For the Maxwell action,
\be
S_{\rm Max}=-\frac1{4g^2}\int_{\mathcal M}d^3x\,F_{\mu\nu}F^{\mu\nu},
\ee
integration by parts gives
\be
\delta S_{\rm Max}
=\frac1{g^2}\int_{\mathcal M}d^3x\,
(\partial_\mu F^{\mu\nu})\delta A_\nu
-\frac1{g^2}\int_{\partial\mathcal M}d^2x\,
n_\mu F^{\mu\nu}\delta A_\nu.
\label{varMax}
\ee
Using $n_\mu=\sig_\alpha\delta_\mu^{\,2}$ gives
Eq.~\eqref{varMax_boundary}.  For the CS term,
\begin{align}
\delta S_{\rm CS}
={}&\kappa\int_{\mathcal M}d^3x\,
\epsilon^{\mu\nu\rho}\partial_\nu A_\rho\,\delta A_\mu
\nn\\
&+\frac\kappa2\int_{\partial\mathcal M}d^2x\,
n_\nu\epsilon^{\mu\nu\rho}A_\mu\delta A_\rho,
\label{varCS}
\end{align}
which gives Eq.~\eqref{varCS_boundary}.  The gauge-fixing variation is
\be
\delta S_{\rm gf}
=\frac1\xi\int_{\mathcal M}d^3x\,
\partial_\mu(\partial\!\cdot\!A)\delta A^\mu
-\frac1\xi\int_{\partial\mathcal M}d^2x\,
n_\mu(\partial\!\cdot\!A)\delta A^\mu.
\label{varGF}
\ee
Because $n_i=0$, its boundary contribution involves only $\delta A^2$.

For constant coefficients, a general tangential one-derivative term is
$C^{ija}A_i\partial_aA_j/2$.  Its part symmetric in $i,j$ is a tangential total
derivative.  In two dimensions the antisymmetric part is
$C^{[ij]a}=d^a\epsilon^{ij}$, proving the completeness of
Eq.~\eqref{Lbd_tan} within the stated truncation.  Its variation is
\be
\delta\int d^2x\,
\frac{d_a}{2}\epsilon^{ij}A_i\partial_aA_j
=\int d^2x\,d_a\epsilon^{ij}\delta A_i\partial_aA_j.
\label{derivative_variation}
\ee
Together with Eqs.~\eqref{varMax_boundary} and~\eqref{varCS_boundary}, this gives
Eqs.~\eqref{BC0_exp}--\eqref{BC1_exp}.

The component signs can be checked explicitly.  Since
$\matE\vectA=(A_1,-A_0)^{\rm T}$, the CS boundary term contributes
$+\sig_\alpha\kappa A_1/2$ to the first equation and
$-\sig_\alpha\kappa A_0/2$ to the second.  The Maxwell contribution is
$-\sig_\alpha F^{2i}/g^2$ on both components.  These signs are fixed by the
outward normal and are not changed by raising the tangential index.

For the normal-field-strength term,
\begin{align}
\delta\int_{\Sigma_\alpha}d^2x\,a_4A_iF_2{}^i
={}&\int_{\Sigma_\alpha}d^2x\,a_4
\left(F_2{}^i\delta A_i+A_i\delta F_2{}^i\right).
\label{a4_variation}
\end{align}
Independent variation of $F_{2i}$ gives Eq.~\eqref{a4_branch_condition}.  If
$a_4\neq0$, it implies $A_i=0$; the coefficient of $\delta A_i$ must still
vanish and, because all tangential terms then vanish on the boundary, gives
Eq.~\eqref{a4_flux_condition}.  Therefore a pure Dirichlet variational branch
occurs only at the tuned value $a_4=\sig_\alpha/g^2$; generic nonzero $a_4$
imposes both $A_i=0$ and $F^{2i}=0$.  This is why the $a_4$ term cannot be
absorbed into the same tangential matrix as Eq.~\eqref{Lbd_tan}.

The gauge variation of the boundary functional is
\be
\delta_\lambda S_{\rm bd}^{\rm tan}
=-\sum_\alpha\int_{\Sigma_\alpha}d^2x\,
\lambda\,\partial_i\mathcal E^i_{(\alpha)},
\label{gauge_var_bd}
\ee
where $\bm{\mathcal E}$ is Eq.~\eqref{boundary_Euler_vector}.  The CS gauge
variation is
\be
\delta_\lambda S_{\rm CS}
=\sum_\alpha\sig_\alpha\frac\kappa2
\int_{\Sigma_\alpha}d^2x\,
\lambda\,\epsilon^{ij}\partial_iA_j.
\label{gauge_var_CS}
\ee
Combining these with $\delta_\lambda S_J=-\int\lambda\partial_iJ^i$ gives
Eq.~\eqref{Ward_W} and identifies the current in
Eq.~\eqref{current_from_Ward}.

For completeness, varying the bulk Gauss generator gives
\be
\delta\left[-\int_{\Sigma_t}d^2x\,\lambda\mathcal G\right]
=\text{bulk terms}
-\sum_\alpha\int_{\Sigma_\alpha\cap\Sigma_t}dx^1\,
\sig_\alpha\lambda
\left(\delta\pi^2-\frac\kappa2\delta A_1\right).
\label{generator_variation}
\ee
The charge Eq.~\eqref{boundary_charge} cancels this term.  On the compatible
family $d_0=d_1=0$, so the boundary functional contributes no additional
symplectic term; the canonical brackets then lead directly to
Eqs.~\eqref{charge_variation}--\eqref{KM_density}.  A branch with $d_0\neq0$
would require a separate canonical analysis including its boundary kinetic
term.

Indeed, $\pi^2=F_{02}/g^2-\kappa A_1/2$, so the combination in
Eq.~\eqref{boundary_charge} is
$F_{02}/g^2-\kappa A_1$.  Under
$\delta_\eta A_1=\partial_1\eta$, the field strength is invariant and the full
variation is $-\kappa\partial_1\eta$.  This fixes the coefficient of the central
term without using a time-ordered two-point identity.

\subsection*{Dirac bracket of the edge scalar}

For completeness, the normalization of Eq.~\eqref{edge_scalar_action} can be
verified directly.  Suppress the boundary label and write
\be
L_{\rm edge}=\frac{1}{2k}\int dx^1\,
\partial_1\phi\left(\partial_0\phi+v\partial_1\phi\right).
\label{edge_L_app}
\ee
The canonical momentum is
\be
\Pi=\frac{1}{2k}\partial_1\phi,
\label{edge_momentum_app}
\ee
so the theory has the second-class constraint
\be
\chi(x)=\Pi(x)-\frac{1}{2k}\partial_1\phi(x)\approx0.
\label{edge_constraint_app}
\ee
On the nonzero-mode sector its Poisson kernel is
\be
\{\chi(x),\chi(y)\}
=-\frac1k\partial_x\delta(x-y),
\label{edge_constraint_kernel}
\ee
whose inverse is $-k\,\mathrm{sgn}(x-y)/2$.  The resulting Dirac brackets are
\be
\{\phi(x),\phi(y)\}_{\rm D}
=-\frac{k}{2}\,\mathrm{sgn}(x-y),
\qquad
\{\partial_1\phi(x),\partial_1\phi(y)\}_{\rm D}
=k\partial_x\delta(x-y).
\label{edge_Dirac_brackets}
\ee
After quantization, the second relation is Eq.~\eqref{scalar_algebra}.  The
Legendre transform gives
\be
H_{\rm edge}=-\frac{v}{2k}\int dx^1\,
(\partial_1\phi)^2,
\label{edge_H_app}
\ee
and its Dirac bracket with $\partial_1\phi$ produces
$(\partial_0+v\partial_1)\partial_1\phi=0$.  The inverse of
Eq.~\eqref{edge_constraint_kernel} is defined only after the constant mode has
been separated, which is the canonical origin of the zero-mode qualification in
Sec.~\ref{edge}.

\section{Derivation of the physical secular equation on the strip}
\label{secular_app}

\subsection*{Compatibility determinant}

For the polarization Eq.~\eqref{physical_polarization}, with
$A_0=-e^0$ and $A_1=e^1$, define
\begin{align}
V_{0,\alpha}(p)
={}&-\frac{\sig_\alpha}{g^2}F^{20}(p)
+b_{00}^{(\alpha)}A_0(p)
+\left[b_{01}^{(\alpha)}+\varrho_\alpha(\omega,k)\right]A_1(p),
\label{V0_app}\\
V_{1,\alpha}(p)
={}&-\frac{\sig_\alpha}{g^2}F^{21}(p)
+\left[b_{01}^{(\alpha)}-\varrho_\alpha(\omega,k)\right]A_0(p)
+b_{11}^{(\alpha)}A_1(p).
\label{V1_app}
\end{align}
The field-strength components needed here are
\be
F^{20}=ip\,e^0-i\omega e^2,
\qquad
F^{21}=ip\,e^1-ik e^2.
\label{F20_F21_app}
\ee
Using $m=\kappa g^2$, substituting Eq.~\eqref{physical_polarization}, and
reducing every occurrence of $\omega^2$ with the mass shell produces a common
kinematic factor.  No division by a boundary coefficient is required, so the
determinant calculation also covers the $b_{00}=0$ charts.  One finds
\be
V_{0,\alpha}(p)V_{1,\alpha}(-p)
-V_{1,\alpha}(p)V_{0,\alpha}(-p)
=\frac{2imp}{m^2+p^2}
\left\{
\det\matB_{(\alpha)}+
[\varrho_\alpha-\sig_\alpha\kappa]^2
\right\},
\label{Delta_app}
\ee
up to an orientation-independent overall phase associated with the normalization
of $e^\mu$.  The terms quadratic and linear in $\omega,k$ vanish for all momenta
only if $d_0=d_1=0$; the remaining constant gives
$\det\matB=-\kappa^2/4$.

The conclusion holds for both orientations because the orientation factor appears
inside a square.  The exceptional kinematic points at which the prefactor in
Eq.~\eqref{Delta_app} vanishes are reached by continuity from generic momentum
and do not generate an additional compatible branch.

To prove the decomposition Eq.~\eqref{on_shell_decomposition}, define
$A_\mu^{\rm phys}=-f_\mu/m$.  The first equation in
Eq.~\eqref{self_dual_equations} gives
$F_{\mu\nu}[A^{\rm phys}]=F_{\mu\nu}[A]$, hence
$A-A^{\rm phys}$ is locally exact.  On the compatible family the physical part
of the boundary equation contains $\matJ\vectA^{\rm phys}$, while the pure-gauge
part contains $\matG\partial\lambda$.  The rank-one identities
Eq.~\eqref{JG_factorization} then prove Eq.~\eqref{physical_gauge_split}.

The complementary $b_{00}=0$, $b_{01}=+\sig\kappa/2$ component has
$\matJ$ with a vanishing first row and hence zero local charge density.  It can
be treated separately, but it does not connect continuously to the finite-density
positive-impedance examples used in the Casimir section.

\subsection*{Gauge factorization and the one-channel determinant}

The rank-one decomposition also fixes the determinant multiplicity.  For fixed
Euclidean tangential momentum, the gauge-invariant field $\psi=f^2$ obeys
\be
(-\partial_2^2+Q^2)\psi=0,
\qquad
Q^2=\zeta^2+k^2+m^2,
\label{normal_scalar_equation_app}
\ee
and therefore has two normal amplitudes.  At either boundary,
Eq.~\eqref{scalar_oblique_BC} supplies one linear relation between the incoming
and outgoing amplitudes.  There is consequently one reflection coefficient per
boundary and one secular function, rather than a determinant over two
independent tangential polarizations.

This conclusion may be seen directly in the multiple-reflection basis.  Let
$A_-$ denote the amplitude arriving at the lower boundary and $A_+$ the amplitude
leaving it.  Reflection and propagation give
\be
A_+=r_0A_-,
\qquad
B_+=e^{-Qh}A_+,
\qquad
B_-=r_hB_+,
\qquad
A_-=e^{-Qh}B_-.
\label{roundtrip_amplitudes_app}
\ee
A nonzero mode exists precisely when
\be
\left(1-r_0r_he^{-2Qh}\right)A_-=0,
\label{one_channel_secular_app}
\ee
which is Eq.~\eqref{secular_equation}.  The factor multiplying $A_-$ is the
complete physical secular determinant at that $(\zeta,k)$.

The residual field $\lambda$ in Eq.~\eqref{on_shell_decomposition} does not
supply a second copy of Eq.~\eqref{normal_scalar_equation_app}.  The
gauge-invariant bulk action vanishes on $A_\mu=\partial_\mu\lambda$, and the
compatible boundary equation reduces independently to
$(\partial_0+v_\alpha\partial_1)\lambda_\alpha=0$ on each component.  After
quotienting by gauge transformations that vanish at both boundaries, the two
boundary values are the local edge variables represented by
Eq.~\eqref{edge_scalar_action}; their nonzero-mode frequencies contain no normal
propagation factor $e^{-Qh}$.

There is also no hidden physical--gauge mixing on the stationary domain.  The
compatible matrix can be written as
\be
\matB_{(\alpha)}=\frac\kappa2
\left(u_\alpha w_\alpha^{\rm T}+w_\alpha u_\alpha^{\rm T}\right).
\label{B_factorization_app}
\ee
Under $A=A^{\rm phys}+d\lambda$, the cross term from the boundary quadratic
functional is $\vectA^{{\rm phys}\,\rm T}\matB\,\partial\lambda$.  The CS bulk
term supplies, after tangential integration by parts, the additional boundary
cross term
$\sig_\alpha\kappa\vectA^{{\rm phys}\,\rm T}\matE\,\partial\lambda/2$.
Their sum is
\be
\vectA^{{\rm phys}\,\rm T}\matG_{(\alpha)}\partial\lambda
=\kappa\big(w_\alpha^{\rm T}\vectA^{\rm phys}\big)
\big(u_\alpha^{\rm T}\partial\lambda\big)=0,
\label{cross_term_cancellation_app}
\ee
where Eq.~\eqref{physical_gauge_split} was used.  Thus the stationary spectrum
is the union of one propagating physical channel and local chiral edge modes.
The latter have the same nonzero-mode spectrum at finite and infinite separation
and cancel from $E_{\rm int}(h)$.  A possible global zero-mode constraint changes
neither the nonzero-momentum secular equation nor the local central terms.  This
is a reduced-phase-space statement; no separate gauge-fixed vector/ghost
determinant is assumed.

\subsection*{Lower boundary, reflection and propagation}

On the compatible family, the first boundary expression evaluated on a physical
wave has numerator
\begin{align}
N_\alpha(p)
={}&i\gamma_\alpha m^2(k-v_\alpha\omega)
+i\sig_\alpha m^2\omega
\nn\\
&+mp\left[
\gamma_\alpha(kv_\alpha-\omega)-\sig_\alpha k
\right].
\label{Nalpha}
\end{align}
The second boundary equation gives the same ratio.  For incident and outgoing
normal momenta $p_{\rm in}=\sig_\alpha iQ$ and
$p_{\rm out}=-\sig_\alpha iQ$,
\be
r_\alpha=-\frac{N_\alpha(p_{\rm in})}{N_\alpha(p_{\rm out})},
\label{r_from_N}
\ee
which reduces to Eq.~\eqref{general_r}.  Propagation from one boundary to the
other multiplies an amplitude by $e^{-Qh}$.  A complete round trip therefore
gives Eq.~\eqref{round_trip_scalar}, and the compatibility condition is
$D=1-M=0$.

For the flip-symmetric family, direct conjugation yields
Eq.~\eqref{conjugate_r}.

The flip acts on the normal momentum as well as on $k$.  The relations
$\gamma_h=\gamma_0$ and $v_h=-v_0$ then turn the upper numerator into the complex
conjugate of the lower one on the Euclidean contour.  This is the step that
reduces the general scalar product $r_0r_h$ to the real reflectivity
Eq.~\eqref{reflectivity}.  The lower amplitude may be written as
\be
r_0=-\frac{
m(\gamma k-ic\zeta)-Q(ck-i\gamma\zeta)}
{m(\gamma k-ic\zeta)+Q(ck-i\gamma\zeta)}.
\label{r0_compact}
\ee
Taking the modulus squared gives Eq.~\eqref{reflectivity}; subtracting numerator
from denominator gives Eq.~\eqref{reflectivity_difference}.

For real $p$, the same numerator satisfies
$N_\alpha(-p)=-N_\alpha(p)^*$, proving $|r_\alpha|=1$.  At imaginary $p$ the
amplitude need not have unit modulus; the sign of $\gamma c$ determines whether
the Euclidean round trip is contractive.

\subsection*{Surface poles and the exceptional chart}

For a lower-boundary mode with $p=i\lambda$, $\lambda>0$, the denominator of the
real-frequency reflection amplitude gives Eq.~\eqref{surface_pole_equation}.

A normalizable mode must satisfy three conditions simultaneously: real
$\omega$, positive $\lambda$, and the physical mass shell.  Squaring the
boundary equation without imposing the sign of $\lambda$ would introduce growing
solutions.  The explicit branch Eq.~\eqref{surface_branch} is obtained before
squaring and therefore keeps the correct sheet.

Eliminating $\lambda$ with the mass shell gives, away from
$\omega^2=k^2$,
\be
(\gamma\omega-ck)^2=m^2(\gamma^2-c^2).
\label{surface_dispersion_factor}
\ee
For $\gamma>c>0$, the decaying solution is Eq.~\eqref{surface_branch}.  It
crosses the singular chart $\omega=k$, $p=im$ at
\be
k_*=m\sqrt{\frac{\gamma+c}{\gamma-c}},
\qquad
\omega_*=k_*,
\qquad
\lambda_*=m.
\label{surface_chart_crossing}
\ee
Solving the first-order helicity equation directly at this point gives the finite
polarization
\be
e_*^\mu\propto
\left(k^2-m^2,\;k^2+m^2,\;2ikm\right).
\label{finite_crossing_polarization}
\ee
Inserting it into Eq.~\eqref{lower_physical_bc} gives
\be
(c-\gamma)k^2+(c+\gamma)m^2=0,
\label{crossing_boundary_condition}
\ee
which is satisfied precisely at Eq.~\eqref{surface_chart_crossing}.  The
complementary light-cone solution $\omega=-k$ at the same normal momentum
$p=im$ has a finite polarization $e^\mu\propto(-1,1,0)$.  Since
$A_0=-e^0$ and $A_1=e^1$, Eq.~\eqref{lower_physical_bc} would then require
$\gamma+c=0$, which has no solution on the positive-impedance component
$\gamma,c>0$.  Thus no physical crossing is missed by using the chart
Eq.~\eqref{physical_polarization} away from $p^2=-m^2$.

At $c=\gamma$, Eq.~\eqref{crossing_boundary_condition} reduces to
$2\gamma m^2=0$, so no finite normalizable mode remains for $m>0$,
$\gamma>0$; equivalently, $k_*\to\infty$ as $c\to\gamma^-$.  If
$c>\gamma>0$, the roots of Eq.~\eqref{surface_dispersion_factor} have normal
momenta with the wrong sign for decay into the lower half-space.  The closure
$\gamma=0$ must be checked separately because the divided dispersion equation is
then degenerate.  The lower condition becomes $A_1=0$; its only apparent decaying
solution from the singular polarization chart is $\omega=k$, $p=im$, but the
finite polarization Eq.~\eqref{finite_crossing_polarization} has
$A_1\propto k^2+m^2\neq0$.  Hence the $\gamma=0$ closure contains no physical
normalizable surface pole.  The upper-boundary result follows by the flip.

\subsection*{Maxwell duality and massless normalization}

At $\kappa=0$, the Maxwell equation and Bianchi identity imply locally
\be
F^{\mu\nu}=g^2\epsilon^{\mu\nu\rho}\partial_\rho\varphi,
\label{Maxwell_duality_app}
\ee
up to an inessential normalization of $\varphi$.  The limiting boundary equations
$F^{20}=F^{21}=0$ become
\be
\partial_1\varphi=0,
\qquad
\partial_0\varphi=0
\qquad\text{on }\Sigma_0,\Sigma_h.
\label{Maxwell_scalar_BC_app}
\ee
For every nonzero tangential mode, this is the Dirichlet condition on the dual
scalar.  The boundary-constant mode is treated by the usual infrared limiting
prescription and does not alter the interaction energy per unit length.

The coefficient in Eq.~\eqref{Maxwell_energy} follows without a polarization
multiplicity:
\begin{align}
E_{\rm int}^{\rm Max}(h)
&=\frac12\int\frac{d^2p}{(2\pi)^2}
\ln(1-e^{-2h|p|})
\nn\\
&=-\frac12\sum_{n=1}^\infty\frac1n
\frac1{2\pi}\int_0^\infty p\,dp\,e^{-2nhp}
=-\frac{\zeta(3)}{16\pi h^2}.
\label{Maxwell_sum_app}
\end{align}
Differentiation gives Eq.~\eqref{Maxwell_force}.  This derivation is independent
of the massive polarization chart and provides a direct check that the physical
strip determinant contains one channel.

\subsection*{Asymptotic integrals}

For generic pole-free data, write
$p_\parallel=m t(\cos\theta,\sin\theta)$.  At threshold,
\be
\Rcal=\Rcal_*\left[1+t^2\mathcal B(\theta)+O(t^4)\right],
\qquad
\Rcal_*=\left(\frac{c-\gamma}{c+\gamma}\right)^2,
\label{generic_R_expansion}
\ee
where
\be
\mathcal B(\theta)=
\frac{c\sin^2\theta-\gamma\cos^2\theta}{c-\gamma}
-\frac{c\sin^2\theta+\gamma\cos^2\theta}{c+\gamma},
\qquad
\int_0^{2\pi}d\theta\,\mathcal B(\theta)=0.
\label{generic_R_angular_average}
\ee
The vanishing angular average implies that the momentum dependence of
$\Rcal$ first contributes beyond the first kinematic power correction.  Keeping
the first round trip through that order therefore gives
\be
E_{\rm int}\simeq-\frac{\Rcal_*}{2}
\int\frac{d^2p_\parallel}{(2\pi)^2}
e^{-2h\sqrt{p_\parallel^2+m^2}}.
\label{first_roundtrip_asymptotic}
\ee
The radial integral is
\be
\int\frac{d^2p_\parallel}{(2\pi)^2}
e^{-2h\sqrt{p_\parallel^2+m^2}}
=\frac{e^{-2mh}}{2\pi}
\left(\frac{m}{2h}+\frac1{4h^2}\right),
\label{radial_asymptotic_integral}
\ee
which gives Eqs.~\eqref{generic_large_E} and~\eqref{generic_large_F}.  On the
transparent line,
\be
\Rcal=\left(\frac{Q-m}{Q+m}\right)^2
=\frac{p_\parallel^4}{16m^4}+O(p_\parallel^6),
\label{transparent_expansion}
\ee
and the Gaussian threshold integral gives
Eq.~\eqref{transparent_asymptotics}.

\end{document}